\documentclass[journal]{IEEEtran}
\usepackage{array,balance}
\usepackage{graphicx}
\usepackage{epsfig}
\usepackage{cite}
\usepackage{subfigure}
\usepackage{epstopdf}
\usepackage{amsfonts,amsmath,amsthm,amssymb,balance}
\usepackage{multirow,balance}
\usepackage{bm}
\usepackage[usenames,dvipsnames]{color}
\usepackage{colortbl}
\usepackage{float}
\usepackage{booktabs}
\usepackage{algorithm}
\usepackage{url}
\usepackage[noend]{algpseudocode}
\definecolor{mygray}{gray}{.9}

\newcolumntype{C}[1]{>{\PreserveBackslash\centering}p{#1}}
\newcolumntype{R}[1]{>{\PreserveBackslash\raggedleft}p{#1}}
\newcolumntype{L}[1]{>{\PreserveBackslash\raggedright}p{#1}}
\usepackage{float}
\begin{document}
\title{Artificial Noisy MIMO Systems under Correlated Scattering Rayleigh Fading -- A Physical Layer Security Approach}
\author{\IEEEauthorblockN{Yiliang Liu, ~\IEEEmembership{Student Member,~IEEE}, Hsiao-Hwa Chen,~\IEEEmembership{Fellow,~IEEE}, \\ Liangmin Wang, ~\IEEEmembership{Member,~IEEE}, and Weixiao Meng,~\IEEEmembership{Senior Member,~IEEE}}
\thanks{Y. Liu (email: {\tt alanliuyiliang@gmail.com}) and W. Meng (email: {\tt wxmeng@hit.edu.cn}) are with the School of Electronics and Information Engineering, Harbin Institute of Technology, China. H.-H. Chen (email: {\tt hshwchen@mail.ncku.edu.tw}) (the corresponding author) is with the Department of Engineering Science, National Cheng Kung University, Taiwan. L. Wang (email: {\tt Jasonwanglm@gmail.com}) is with the Department of Internet of Things Engineering, Jiangsu University, China.}
\thanks{This work was supported in part by National Natural Science Foundation of China (Nos. U1764263 and 61671186) and Taiwan Ministry of Science and Technology (Nos. 106-2221-E-006-028-MY3 and 106-2221-E-006-021-MY3).}
\thanks{The paper was submitted on November 23, 2018, and revised on \today.}}
\markboth{Accepted in IEEE Systems Journal, YEAR 2019}{Liu \MakeLowercase{\textit{et al.}}: Artificial Noisy MIMO Systems\ldots}

\maketitle

\begin{abstract}
The existing investigations on artificial noise (AN) security systems assumed that only null spaces is used to send AN signals, and all eigen-subchannels should be used to transmit messages. Our previous work proposed an AN scheme that allocates some of eigen-subchannels to transmit AN signals for improving secrecy rates. Nevertheless, our previous work considered only uncorrelated MIMO Rayleigh fading channels. In fact, the correlations among antennas exist in realistic scattering channel environments. In this paper, we extend our previous AN scheme to spatially correlated Rayleigh fading channels at both legitimate receiver- and eavesdropper-sides and derive an exact theoretical expression for the ergodic secrecy rate of the AN scheme, along with an approximate analysis. Both numerical and simulation results show that the proposed AN scheme offers a higher ergodic secrecy rate than the existing schemes, revealing a fact that the correlation among eavesdropper's antennas can potentially improve the secrecy rate of an MIMO system. 
\end{abstract}

\IEEEpeerreviewmaketitle

\section{INTRODUCTION}
Physical layer security has attracted a lot of attention due to its potential to offer low-cost and high-level security in wireless communications\cite{WangHeter2018,Oggier2008,LiuMIMO2017,Li2016, Yu2018sj}. The idea of utilizing artificial noise (AN) or jamming signals, as a physical layer security scheme, was proposed for the first time in Negi and Goel's work \cite{Negi05,Goel2008}. Recently, AN schemes have been extended to different channels to safeguard sensitive and confidential data \cite{Tsai2014, Liu2015, Yun2017conf,Yun2018}. For instance, \cite{Tsai2014, Liu2015, Yun2017conf} considered Rayleigh MIMO channels, whereas \cite{Yun2018} assumed Rician MIMO channels. The basic idea of the aforementioned AN schemes is that message streams are sent in a multiplex mode via all eigen-subchannels (positive eigenvalue channels) at desired directions, and the AN signals are transmitted to a null space of desired directions, such that they do not interfere desired users but only impair eavesdropped channels.

However, these AN schemes use a null space for AN signals only under the condition that the number of transmit antennas is larger than that of receivers \cite{Negi05,Goel2008, Wang2017AN, Tsai2014, Liu2015, Yun2017conf,Yun2018,Shu2018}. In addition, using all eigen-subchannels in an MIMO system for message transmission may degrade secrecy rate if compared to the schemes, which properly allocate some of eigen-subchannels for AN signals. In our previous work \cite{Liuwishart2017}, we took the number of eigen-subchannels of message streams as a variable that can be leveraged to maximize ergodic secrecy rate and showed that, when the number of transmit antennas is smaller than that of receivers, it is possible to find eigen-subchannels used by AN signals in order not to interfere desired users with the help of AN elimination technique at the desired users. The work in \cite{Liuwishart2017} was done based on a Rayleigh fading channel, as Rayleigh fading is a reasonable model for heavily built-up urban environments \cite{Wang2017AN}, which has been extended to uncorrelated Rician fading channels via a non-central Wishart matrix in \cite{Ahmed2018}. Zheng \textsl{et al.} also used eigen-subchannels for AN signal transmission, but treated the AN as interference signals due to the lack of proper AN elimination techniques \cite{Zheng2018}. 

It is noted that all of the aforementioned schemes assumed the presence of uncorrelated fading in MIMO channels. Unfortunately, in many real applications, the correlation among antennas may exist due to poor-scattering environments or small spacing between antenna elements \cite{3gpp, Simon2006,Bolcskei2003}. It motivates us to design a better AN scheme to suit for correlated fading environments. Recently, Li \cite{Li2018} investigated secure transmissions in an MISO-based system with receiver-side correlation in satellite-terrestrial channels. The effect of double-side correlation in the main and wiretap channels of MIMO systems was studied via Monte Carlo simulations in \cite{Zorgui2016} and \cite{Zhang2016}. All of the above investigations showed that the correlation has its impacts on security performance. However, the effect of receiver-side correlation of MIMO-aided AN systems has not been fully investigated so far, and an exact expression for ergodic secrecy rate of AN schemes is far more useful than Monte Carlo simulation results because it provides us an objective function to disclose the relationship between secrecy rate and channel correlation.

This paper focuses on receiver-side correlated fading scenarios at both legitimate receiver- and eavesdropper-sides. As shown in a report on the downlink channel correlation by 3GPP \cite{3gpp}, transmitters are located at base stations with enough space to deploy multiple antennas, and the size of a receiver (e.g., a mobile terminal) is usually small. Thus, receiver-side correlation more likely occurs than transmitter-side correlation in downlink channels. In addition, in 5G and beyond systems, the receivers, such as vehicles and unmanned aerial vehicles (UAVs), may move to an appropriate location for secrecy transmission, whose channel correlation parameters may change from time to time based on statistical channel information \cite{Lyu2018}, such as mean angles of arrival (AoA) and receive angle spread (RAS). Some devices with a very small antenna separation distance (such as massive MIMO) will emerge for secure communications in the future.

The main contributions of this work can be summarized as follows.
\begin{enumerate}
\item We extend the AN scheme \cite{Liuwishart2017} to receiver-side correlated MIMO channels, and derive an exact expression for ergodic secrecy rates. To the best of our knowledge, this is the first time to give such an exact expression in terms of spatial correlation parameters (i.e., mean AoA, RAS, and antenna spacing, etc.). A suitable number of eigen-subchannels for messages and AN can be easily identified based on the derived ergodic secrecy rate expression. Then, we simplify the expression and give its approximate analysis.
\item In addition, we derive an exact closed-form expression for marginal probability density function (pdf) of the $k$th eigenvalue of receiver-side correlated Wishart matrices. The work in \cite{Ordonez2009} required two expressions to formulate this function. We need only one expression as a more generalized form. We identify the properties of the correlated matrices in terms of spatial correlation parameters. The mathematical investigations given in the paper are general, which can also be used for analyzing ergodic secrecy rates of an AN scheme and channel capacities of traditional MIMO systems.
\end{enumerate}

The remainder of this paper can be outlined as follows. Section II introduces the system model and AN scheme. Section III aims to derive an exact mathematical expression for ergodic secrecy rates, along with an approximate analysis. Section IV is dedicated for numerical analysis and simulations, followed by the conclusions in Section V.

The notations are explained as follows. Bold uppercase letters denote matrices and bold lowercase letters denote column vectors. $\mathbf{A}^{\dagger}$ represents the Hermitian transpose of $\mathbf{A}$. $\mathbf{I}_a$ is an identity matrix with its rank $a$. $\mathbf{S}_a$ denotes an $(a \times a)$ square matrix with its order $a$. E$[\cdot]$ denotes the expectation operator. $[\mathbf{A}]_{i,j}$ gives the $i$th row and the $j$th column element of $\mathbf{A}$. $[\mathbf{A}]_{(i\sim u),(j\sim v)}$ is a submatrix of $\mathbf{A}$, including the $i$th to the $u$th rows and the $j$th to the $v$th columns of $\mathbf{A}$. $\exp(x)$ denotes an exponential function of $x$. $\det[\mathbf{A}]$ is the determinant of $\mathbf{A}$. $\text{etr}(\mathbf{X})$ denotes $\exp[\text{Tr}(\mathbf{X})]$, where $\text{Tr}(\mathbf{X})$ is the trace of $\mathbf{X}$. $\otimes$ stands for a Kronecker product. An $[a\times(b+c)]$ matrix $[\mathbf{A},\mathbf{B}]$ denotes a combined matrix between an $(a\times b)$ matrix $\mathbf{A}$ and an $(a\times c)$ matrix $\mathbf{B}$. $(\mathbf{A})^{1/2}$ represents matrix square root operation such that $\mathbf{A}^{1/2}(\mathbf{A}^{1/2})^{\dagger}=\mathbf{A}$. $\binom{x}{y}$ is the combination between $x$ and $y$ such that $\binom{x}{y}=\frac{x!}{(x-y)!y!}$.

\begin{figure}[h]
\centering
\includegraphics[width=1\linewidth]{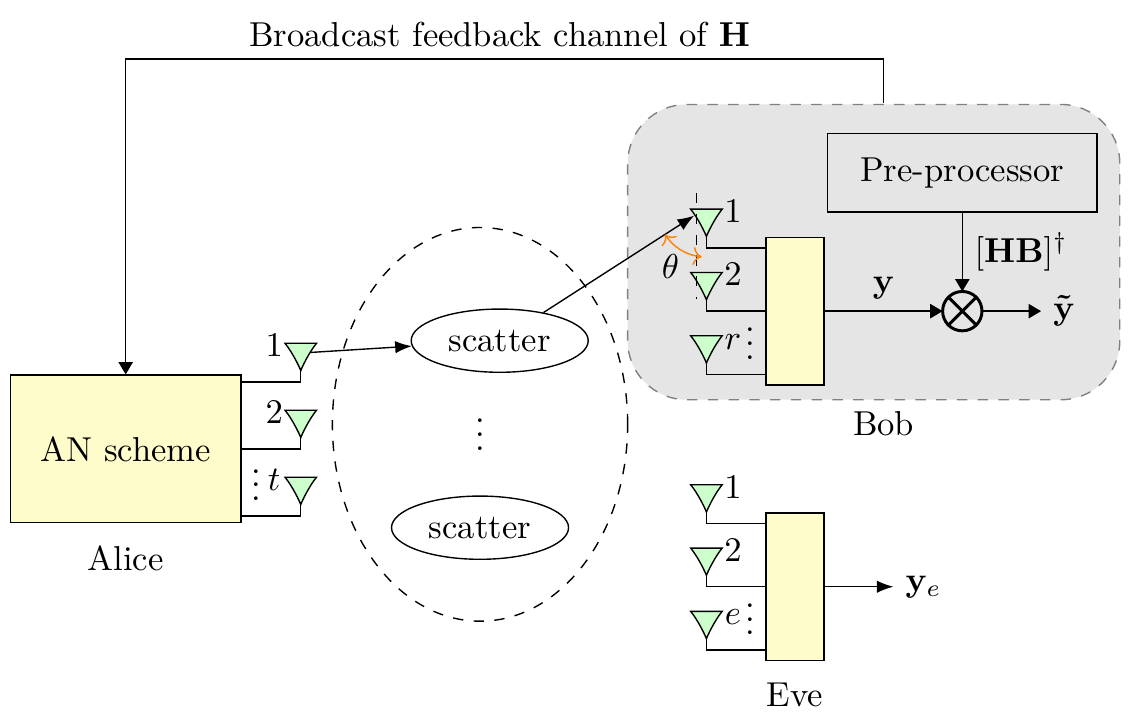}
\caption{Illustration of an artificial noisy MIMO wiretap channel model, where Alice, Bob, and Eve use uniformly linear array antennas, and $\theta$ is AoA between a scattered path and the antenna array.}\label{model}
\end{figure}

\section{SYSTEM MODEL}

In this section, we introduce a system model that specifies a spatial correlation channel, as well as the AN scheme.

\subsection{MIMO Wiretap Channel with Spatial Correlation}
Let us consider an MIMO communication system in the presence of correlated Rayleigh fading at both legitimate receiver- and the eavesdropper-sides. The system consists of a transmitter (Alice) with $t$ transmit antennas, a legitimate receiver (Bob) with $r$ receive antennas, and an eavesdropper (Eve) with $e$ receive antennas, as shown in Fig. \ref{model}, where $t>e$ and $r$ is arbitrary. In general, the main channel between Alice and Bob and the wiretap channel between Alice and Eve are defined by receiver-side correlated complex Gaussian matrices $\mathbf{H}\in\mathbb{C}^{r\times t}$ and $\mathbf{H}_e\in\mathbb{C}^{e\times t}$, as given in Definition 1. $\mathbf{R}_r\in\mathbb{C}^{r\times r}$ and $\mathbf{R}_e\in\mathbb{C}^{e\times e}$ are the receiver-side correlated channel matrices of Bob and Eve, respectively, as given in Definition 2. 

\vspace{0.1in}
\textsl{Definition 1} (Central complex Gaussian matrix): Each element of a random matrix $\mathbf{A}\in\mathbb{C}^{a\times b}$ takes a complex value, whose real and imaginary parts follow a normal distribution $\mathcal{N}(0, 1/2$). $\mathbf{A}$ is defined as a central complex Gaussian matrix with a covariance matrix $\mathbf{\Phi}_a\otimes \mathbf{\Psi}_b$, which is expressed as
\begin{equation}
\mathbf{A}\sim \mathcal{CN}_{a,b}(\mathbf{0},\mathbf{\Phi}_a\otimes \mathbf{\Psi}_b),
\end{equation}
where $\mathbf{\Psi}_b=\text{E}[\mathbf{a}_{i,(1\sim b)}\mathbf{a}_{i,(1\sim b)}^{\dagger}]$ for $i=1, ..., a$, and $\mathbf{\Phi}_a=\text{E}[\mathbf{a}_{(1\sim a),j}\mathbf{a}_{(1\sim a),j}^{\dagger}]$ for $j=1, ..., b$. The $(a\times a)$ matrix $\mathbf{\Phi}_a$ and $(b\times b)$ matrix $\mathbf{\Psi}_b$ are the Hermitian positive definite matrices. A similar definition of this complex Gaussian matrix can be found in \cite{Ordonez2009,Ratnarajah2003}. 

\vspace{0.1in}
In order to investigate $\mathbf{H}$ and $\mathbf{H}_e$ in the model, let us use a Kronecker model to define 
\begin{equation}
\mathbf{H}=\mathbf{R}_r^{1/2}\mathbf{H}_{\text{Bob}} \sim \mathcal{CN}_{r,t}(\mathbf{0},\mathbf{R}_r\otimes \mathbf{I}_t),
\end{equation}
\begin{equation}
\mathbf{H}_e=\mathbf{R}_e^{1/2}\mathbf{H}_{\text{Eve}} \sim \mathcal{CN}_{e,t}(\mathbf{0},\mathbf{R}_e\otimes \mathbf{I}_t),
\end{equation}
where $\mathbf{H}_{\text{Bob}}\in\mathbb{C}^{r\times t}$ and $\mathbf{H}_{\text{Eve}}\in\mathbb{C}^{e\times t}$ are complex Gaussian random matrices with independent complex Gaussian elements. Similar to $\mathbf{A}$, the real and imaginary parts of each element of $\mathbf{H}_{\text{Bob}}$ and $\mathbf{H}_{\text{Eve}}$ follow a normal distribution $\mathcal{N}(0, 1/2$). $\mathbf{H}_{\text{Bob}}$ and $\mathbf{H}_{\text{Eve}}$ can be expressed respectively as
\begin{eqnarray}
&&\mathbf{H}_{\text{Bob}}\sim \mathcal{CN}_{r,t}(\mathbf{0},\mathbf{I}_r \otimes \mathbf{I}_t)\label{hu1},\\
&&\mathbf{H}_{\text{Eve}}\sim \mathcal{CN}_{e,t}(\mathbf{0},\mathbf{I}_e \otimes \mathbf{I}_t)\label{hu2}.
\end{eqnarray}

The correlated matrices $\mathbf{R}_r$ and $\mathbf{R}_e$ are the key factors in deriving channel state information (CSI) matrices. From \cite{McKay2005,Bolcskei2003}, we know that a correlated matrix $\mathbf{R}_a$ (a generalized version of $\mathbf{R}_r$ and $\mathbf{R}_e$) is a function of AoA distribution (defined by $\theta$), as given in Definition 2, which is a way to generate a receiver-side correlated matrix.

\vspace{0.1in}
\textsl{Definition 2} (Receiver-side correlated matrix): Assume that all antennas form a uniformly linear antenna array with $d=d_{\min}/\omega$, where $d$ is the normalized minimum distance, $d_{\min}$ is the spacing between any two neighbor antennas, and $\omega$ is the wavelength. Each element of a receiver-side correlated matrix $\mathbf{R}_a$, i.e., $[\mathbf{R}_a]_{u,v}$ is  
\begin{flalign}\label{corrematrix}
[\mathbf{R}_a]_{u,v}&=\exp\big\{-j2\pi d(u-v)\cos\bar{\theta}\big\} \\ \notag
&\times \exp\big\{-\frac{1}{2}\big[2\pi d \delta (u-v)\sin\bar{\theta}\big]^2\big\},	
\end{flalign}
where $u\in\{1,...,a\}$ and $v\in\{1,...,a\}$ are the receive antenna index numbers. For Bob, we have $a=r$, and for Eve, we have $a=e$. The AoA, i.e., $\theta$, follows a Gaussian distribution, where the mean AoA of $\theta$ is $\bar{\theta}$ and the RAS (variance) of $\theta$ is $\delta$. A similar definition of this receiver-side correlated matrix can be found in \cite[Eqn. (4)]{Bolcskei2003} and \cite[Eqn. (106)]{McKay2005}. Based on the calculations of Eqn. (\ref{corrematrix}), we can see that when $u=v$, $\big|[\mathbf{R}_a]_{u,v}\big|$ equals to one. When $u\neq v$, $\big|[\mathbf{R}_a]_{u,v}\big|$ approaches to zero with an increasing $d$, $\bar{\theta}$, or $\delta$. Hence, $\mathbf{R}_a$ approaches to $\mathbf{I}_a$ with an increasing $d$, $\bar{\theta}$, or $\delta$. This means that the correlation will be reduced with an increasing $d$, $\bar{\theta}$, or $\delta$.

\vspace{0.1in}
Let us use Theorem 1 to specify the properties of the correlated matrix $\mathbf{R}_a$, which is useful for the approximate analysis of ergodic secrecy rates in the next section.

\vspace{0.1in}
\textsl{Theorem 1}: Let $\mathbf{R}_a(d)$, $\mathbf{R}_a(\bar{\theta})$, and $\mathbf{R}_a(\delta)$ be the functions of $d$, $\bar{\theta}$, and $\delta$, respectively, as given in Definition 2. The largest eigenvalue of $\mathbf{R}_a$ is defined as $\sigma_1(\mathbf{R}_a)$, and the determinant of $\mathbf{R}_a$ is defined as $\det[\mathbf{R}_a]$. Then, we get the conclusions as follows.
\begin{itemize}
\item 
If $d_1>d_2$, we have $\sigma_1\big[\mathbf{R}_a(d_1)\big]<\sigma_1\big[\mathbf{R}_a(d_2)\big]$ and $\det[\mathbf{R}_a(d_1)]>\det[\mathbf{R}_a(d_2)]$.  
\item 
If $\bar{\theta}_1>\bar{\theta}_2$, we have $\sigma_1\big[\mathbf{R}_a(\bar{\theta}_1)\big]<\sigma_1\big[\mathbf{R}_a(\bar{\theta}_2)\big]$ and $\det[\mathbf{R}_a(\bar{\theta}_1)]>\det[\mathbf{R}_a(\bar{\theta}_2)]$.
\item 
If $\delta_1>\delta_2$, we have $\sigma_1\big[\mathbf{R}_a(\delta_1)\big]<\sigma_1\big[\mathbf{R}_a(\delta_2)\big]$ and $\det[\mathbf{R}_a(\delta_1)]>\det[\mathbf{R}_a(\delta_2)]$.
\end{itemize}

\textsl{Proof:} See Appendix A. 

\vspace{0.1in}
Let us define the mean AoAs at Bob and Eve as $\bar{\theta}_{\text{Bob}}$ and $\bar{\theta}_{\text{Eve}}$, respectively, define RASs at Bob and Eve as $\delta_{\text{Bob}}$ and $\delta_{\text{Eve}}$, respectively, and define the normalized distances at Bob and Eve as $d_{\text{Bob}}$ and $d_{\text{Eve}}$, respectively.     

\subsection{Artificial Noise Precoding}
In this paper, we use the AN scheme as proposed in \cite{Liuwishart2017}. There are $s_1$ eigen-subchannels for sending confidential messages selected by Alice based on CSI feedback from Bob. $s_1$ is a variable that can be adjusted by Alice. More specifically, Alice performs the eigenvalue decomposition (eig) of $\mathbf{H}^{\dagger}\mathbf{H}$, which outputs two unitary matrices, i.e., $\mathbf{U}\in\mathbb{C}^{t\times t}$ and its Hermitian transpose $\mathbf{U}^{\dagger}\in\mathbb{C}^{t\times t}$. The eig process also outputs a diagonal matrix $\mathbf{\Lambda}\in\mathbb{R}^{t\times t}$, which consists of the positive and zero eigenvalues of $\mathbf{H}^{\dagger}\mathbf{H}$, i.e., $(\lambda_1,...,\lambda_t)$, where the positive eigenvalues are defined as $\lambda_1>...>\lambda_n$, where $n=\min(t,r)$. 

Alice generates a message precoding matrix $\mathbf{B} \in \mathbb{C}^{t \times s_1}$, whose columns are the eigenvectors corresponding to the first to the $s_1$th largest eigenvalues of $\mathbf{H}^{\dagger}\mathbf{H}$, and an AN precoding matrix $\mathbf{Z} \in \mathbb{C}^{ t\times s_2}$ $(s_1+s_2=t)$, whose columns are the eigenvectors of the remaining eigenvalues of $\mathbf{H}^{\dagger}\mathbf{H}$. 

\vspace{0.1in}
\textsl{Remark 1:} (Proved in \cite[Le. 1]{Liuwishart2017}): We can readily show $[\mathbf{H}\mathbf{B}]^{\dagger}\mathbf{HZ}=\mathbf{0}$, $[\mathbf{H}\mathbf{B}]^{\dagger}\mathbf{H}_e\mathbf{Z}\neq\mathbf{0}$, and $[\mathbf{H}_e\mathbf{B}]^{\dagger}\mathbf{H}_e\mathbf{Z}\neq\mathbf{0}$.

\vspace{0.1in}
As the CSI is extremely important in this work, we would like to discuss about the CSI at Alice and Eve as follows. 
\begin{itemize}
\item
CSI at Alice: As mentioned earlier, let us consider a slow-fading environment that Alice knows full CSI of Bob, including $\mathbf{H}$ and $\mathbf{R}_r$, via a unprotected broadcast feedback channel from Bob due to FDD or non-reciprocal TDD systems \cite{Kobayashi2011}, but knows only $\mathbf{R}_e$ and the channel distribution information (CDI) of Eve. Alice can get the knowledge of $\mathbf{R}_e$ and the CDI of Eve, because Eve can be just a normal receiver in the same communication system with Alice, and may exchange messages without security protection. Hence, Alice can obtain $\mathbf{R}_e$ via historical CSI of $\mathbf{H}_e$, i.e., $\mathbf{R}_e=\text{E}(\mathbf{H}_e\mathbf{H}_e^{\dagger}/t)$ or statistical AoA information as shown in Definition 2. Otherwise, Alice should assume that there is no correlation at Eve side, i.e., $\mathbf{R}_e=\mathbf{I}_e$, which is the worst assumption because $\mathbf{R}_e=\mathbf{I}_e$ will maximize the ergodic wiretap channel capacity among all realizations of $\mathbf{R}_e$ \cite{Simon2006}. 
\item
CSI at Eve: Let us consider a pessimistic scenario that Eve knows the CSI of all channels, includes $\mathbf{H}$, $\mathbf{H}_e$, $\mathbf{R}_r$, and $\mathbf{R}_e$. This scenario usually exists in feedback-based CSI estimation. The investigation in \cite{Liuma2015} provided an example of the leaked CSI, where Alice sends a training signal to Bob and Bob uses feedback channels to inform Alice of CSI, which allows Bob and Alice to obtain accurate knowledge of $\mathbf{H}$. However, Eve can obtain $\mathbf{H}$ due to the broadcasting nature of feedback channels, and Eve can intercept the training signals to get $\mathbf{H}_e$. In addition, Eve can obtain $\mathbf{R}_r=\text{E}(\mathbf{H}\mathbf{H}^{\dagger}/t)$ and $\mathbf{R}_e=\text{E}(\mathbf{H}_e\mathbf{H}_e^{\dagger}/t)$ according to long-term realizations of $\mathbf{H}$ and $\mathbf{H}_e$ or statistical AoA information as shown in Definition 2.
\end{itemize}

Based on the precoding of $\mathbf{B}$ and $\mathbf{Z}$, Alice transmits a combined signal $\mathbf{w}$ via $t$ antennas as $\mathbf{w}=\mathbf{Bx}+\mathbf{Zv}$, and the received signals at Bob and Eve can be expressed as
\begin{flalign}
&\mathbf{y}=\mathbf{HBx}+\mathbf{H}\mathbf{Zv}+\mathbf{n}\label{q1:subeq1},\\
& \mathbf{y}_e=\mathbf{H}_e\mathbf{Bx}+\mathbf{H}_e\mathbf{Zv}+\mathbf{n}_e, \label{q1:subeq2}
\end{flalign}
respectively. Here, $\mathbf{x}$ is a transmit signal of the desired user, and $\mathbf{v}$ is an AN signal. We follow a convention used in \cite{Negi05,Goel2008}, which used Gaussian input alphabets and Gaussian AN, i.e., both $\mathbf{x}$ and $\mathbf{v}$ are circularly symmetric complex Gaussian vectors with zero-means and covariance matrices $P/t\mathbf{I}_{s_1}$ and $P/t\mathbf{I}_{s_2}$, respectively, where $P$ is an average transmit power constraint. For analytical simplicity, we distribute total power over all antennas equally as $\rho=P/t$. $\mathbf{n}$ and $\mathbf{n}_e$ are the additive white Gaussian noise (AWGN) vectors with their covariance matrices $\mathbf{I}_r$ and $\mathbf{I}_e$, respectively.

It is obvious that each antenna transmits a combination of message and AN components, but the AN components can be eliminated by the pre-processor at Bob, who eliminates the AN signal $\mathbf{v}$ by pre-processing ($[\mathbf{H}\mathbf{B}]^{\dagger}\mathbf{HZ}=\mathbf{0}$), and the received signal $\mathbf{y}$ is
\begin{equation}\label{AES1}
\mathbf{\tilde{y}}=[\mathbf{H}\mathbf{B}]^{\dagger}\mathbf{y}=
\mathbf{\Lambda}_{s_1}\mathbf{x}+\mathbf{\tilde{n}},
\end{equation}
where $\mathbf{\tilde{n}}=[\mathbf{H}\mathbf{B}]^{\dagger}\mathbf{n} \in \mathbb{C}^{s_1\times1}$ is an AWGN vector with its distribution $\mathcal{CN}(\mathbf{0},\mathbf{\Lambda}_{s_1})$. $\mathbf{\Lambda}_{s_1}\in \mathbb{R}^{s_1\times s_1}$ is a diagonal matrix formed by the first to the $s_1$th eigenvalues of $\mathbf{H}^{\dagger}\mathbf{H}$. In the AN elimination process, the channel, where the received signal is left-multiplied by a given matrix $[\mathbf{H}\mathbf{B}]^{\dagger}$, will not change its capacity if $\mathbf{B}$ includes all eigenvectors of $\mathbf{H}^{\dagger}\mathbf{H}$. Since we have $[\mathbf{H}\mathbf{B}]^{\dagger}\mathbf{H}_e\mathbf{Z}\neq\mathbf{0}$ and $[\mathbf{H}_e\mathbf{B}]^{\dagger}\mathbf{H}_e\mathbf{Z}\neq\mathbf{0}$, Eve can not eliminate this AN signal under the condition of $t>e$, such that the AN signal degrades Eve's channel capacity even if Eve has the knowledge of $\mathbf{H}$, $\mathbf{H}_e$, $\mathbf{B}$, and $\mathbf{Z}$. In this way, we can enlarge the capacity difference between the main and wiretap channels. 

\vspace{0.1in}
\section{EXACT AND APPROXIMATE ERGODIC SECRECY RATES}
Next, we derive an exact ergodic secrecy rate expression, as well as perform an approximate analysis to show the impacts of correlated matrices on the ergodic secrecy rates.  

\subsection{Exact Expression for Ergodic Secrecy Rate}
In the proposed scheme, $P$, $\mathbf{H}$, $\mathbf{R}_r$, and $\mathbf{R}_e$ are system parameters. The numbers of message and AN streams, denoted by $s_1$ and $s_2$, are the variables controlled by us. Then, we can get a real ergodic secrecy rate expression $\tilde{R}_s$ as
\begin{flalign}\label{ISC1}
&\tilde{R}_s(P,\mathbf{H},\mathbf{R}_r, \mathbf{R}_e;s_1,s_2)=\text{E}_{\mathbf{H}_e,\mathbf{H}}[C_m-C_w]^+\\ \notag 
&\geq \big[\text{E}_\mathbf{H}[C_m]-\text{E}_{\mathbf{H}_e,\mathbf{H}}[C_w]\big]^{+}, \label{inequa}
\end{flalign}
where we have $[x]^+=\max(x,0)$, and
\begin{flalign}
C_m=&\log_2\det(\mathbf{I}_r+\rho\mathbf{H}_1\mathbf{H}_1^{\dagger}), \\ \notag
C_w=&\log_2\text{det}\bigg(\mathbf{I}_e+\frac{\rho\mathbf{H}_2\mathbf{H}_2^{\dagger}}{\rho\mathbf{H}_3\mathbf{H}_3^{\dagger}+\mathbf{I}_e}\bigg) \\  
=&\log_2\text{det}\big(\mathbf{I}_e+\rho\mathbf{H}_4\mathbf{H}_4^{\dagger}\big)-\log_2\text{det}\big(\mathbf{I}_e+\rho\mathbf{H}_3\mathbf{H}_3^{\dagger}\big) \label{cw2}.
\end{flalign}
Here, we have $\mathbf{H}_1=\mathbf{HB}\in\mathbb{C}^{r\times s_1}$, $\mathbf{H}_2=\mathbf{H}_e\mathbf{B}\in\mathbb{C}^{e\times s_1}$, $\mathbf{H}_3=\mathbf{H}_e\mathbf{Z}\in\mathbb{C}^{e\times s_2}$, and $\mathbf{H}_4=[\mathbf{H}_2,\mathbf{H}_3]=\mathbf{H}_e\mathbf{U}\in\mathbb{C}^{e\times t}$. 

Note that $C_m$ is the main channel capacity that can be achieved by the pre-processor as shown in Eqn. (\ref{AES1}) \cite{Liuwishart2017}. The pre-processor can eliminate the interference among antennas and AN-induced interference, so that Bob can decode confidential message streams individually. Assume that Eve sees the Gaussian AN signal and AWGN as a combined AWGN, views $\mathbf{H}_2$ as its CSI, and then uses the minimum mean squared error (MMSE) with successive interference cancellation (SIC) technique based on $\mathbf{H}_2$ to achieve a wiretap channel capacity, i.e., $C_w$. From the conclusions made in \cite{Liu2015} and \cite[Ch. 8]{tse2005fundamentals}, the MMSE with SIC technique is the best choice for Eve without knowledge of Gaussian AN signals.

We have an equality in Eqn. (\ref{inequa}) if and only if the secrecy rates are always nonnegative over all channel states. With a large $s_2$, i.e., more eigen-subchannels are allocated for sending AN signals, $C_m$ is much larger than $C_w$ with a high probability\footnote{The test results are available in \url{https://github.com/yiliangliu1990/liugit_pub}.}. However, due to the lack of the knowledge of $\mathbf{H}_e$, we can not determine if an instantaneous secrecy rate is nonnegative or not, and thus we resort to derive a lower bound of the real ergodic secrecy rate as
\begin{equation}\label{lowesc}
R_s(P,\mathbf{H},\mathbf{R}_r, \mathbf{R}_e;s_1,s_2)=[\text{E}_\mathbf{H}[C_m]-\text{E}_{\mathbf{H}_e,\mathbf{H}}[C_w]]^{+},
\end{equation}
assuming that both $\mathbf{H}$ and $\mathbf{H}_e$ are independent receiver-side correlated complex Gaussian matrices. 

In order to calculate the ergodic secrecy rate, we need to calculate $\text{E}_{\mathbf{H}_e,\mathbf{H}}[C_w]$, and we should find out the distributions of random matrices $\mathbf{H}_2$, $\mathbf{H}_3$, and $\mathbf{H}_4$, all of which are the product of a complex Gaussian matrix and an independent unitary matrix. The corresponding results are given in Theorem 2. 

\vspace{0.1in}
\textsl{Theorem 2}: Define $\mathbf{H}_e\sim \mathcal{CN}_{e,t}(\mathbf{0},\mathbf{R}_e\otimes \mathbf{I}_t)$ as a receiver-side correlated central complex Gaussian matrix, and establish an independent $(t\times f)$ unitary matrix $\mathbf{F}$ (generalized for $\mathbf{B}$ and $\mathbf{Z}$). We have
\begin{equation}
\mathbf{H}_e\mathbf{F}\sim \mathcal{CN}_{e,f}(\mathbf{0},\mathbf{R}_e\otimes \mathbf{I}_f),
\end{equation}
where $f\in \mathbb{N}$ and $t\geq f$.

\vspace{0.1in}
\textsl{Proof:} See Appendix B. 

\vspace{0.1in}
From Theorem 2, we know that $\mathbf{H}_2$, $\mathbf{H}_3$, and $\mathbf{H}_4$ are complex Gaussian matrices with their distributions as
\begin{flalign}
&\mathbf{H}_2=\mathbf{H}_e\mathbf{B}\sim \mathcal{CN}_{e,s_1}(\mathbf{0},\mathbf{R}_e\otimes \mathbf{I}_{s_1}),\\
&\mathbf{H}_3=\mathbf{H}_e\mathbf{Z}\sim \mathcal{CN}_{e,s_2}(\mathbf{0},\mathbf{R}_e\otimes \mathbf{I}_{s_2}),\\
&\mathbf{H}_4=\mathbf{H}_e\mathbf{U}\sim \mathcal{CN}_{e,t}(\mathbf{0},\mathbf{R}_e\otimes \mathbf{I}_t),
\end{flalign}
respectively. In order to evaluate the performance of the AN scheme further, we should use the pdf of the $k$th eigenvalue of complex Wishart matrices to derive a theoretical ergodic secrecy rate expression of Eqn. (\ref{lowesc}). Here, we give the definition of the Wishart matrix, as shown in Definition 3.

\vspace{0.1in}
\textsl{Definition 3} (Receiver-side correlated central complex Wishart matrix): For $\mathbf{A}\sim \mathcal{CN}_{a,b}(\mathbf{0},\mathbf{R}_a\otimes \mathbf{I}_b)$, $m=\max(a,b)$, and $n=\min(a,b)$, a Hermitian matrix $\mathbf{W}\in\mathbb{C}^{n\times n}$ is defined as
\begin{equation}\label{WM}
\mathbf{W}=
\begin{cases}
\mathbf{A}\mathbf{A}^{\dagger},&\mbox{$b\geq a$},\\
\mathbf{A}^{\dagger}\mathbf{A}, &\mbox{$b<a$},
\end{cases}
\end{equation}
where $\mathbf{W}$ is called a receiver-side correlated central Wishart matrix defined as $\mathbf{W}\sim W_n(m,\mathbf{0}_n,\mathbf{R}_a)$ with $n$ degrees of freedom, and a receiver-side correlated matrix $\mathbf{R}_a$ has its eigenvalues $\sigma_i, 1\leq i\leq a$, where $\sigma_1>\sigma_2>...>\sigma_a$. The Wishart matrix was investigated first in \cite{wishart1928generalised}.

\vspace{0.1in}
An arbitrary MIMO channel ($\mathbf{H}$, $\mathbf{H}_3$, or $\mathbf{H}_4$) can be effectively decomposed into multiple parallel SISO eigen-subchannels. With the help of transmit and receive signal processing as described in Section II, $s_1$ eigen-subchannels are selected for sending messages. Then, we can re-write the ergodic secrecy rate function Eqn. (\ref{lowesc}) as
\begin{flalign}\label{esc}
&R_s(P,\mathbf{R}_r, \mathbf{R}_e;s_1,s_2)\\ \notag
&=\big[C_{\mathbf{H}}(\mathbf{R}_r, \rho,s_1)+C_{\mathbf{H}_3}(\mathbf{R}_e, \rho,n_1)-C_{\mathbf{H}_4}(\mathbf{R}_e,\rho,e)\big]^{+},
\end{flalign}
where 
\begin{flalign}\label{esc1}
C_{\mathbf{A}}(\mathbf{R}_a, \rho,\eta)=\sum_{k=1}^{\eta}\int_0^{\infty}\log_2(1+\rho x)f_{\lambda_k}(x)dx,
\end{flalign}
in which we have $\rho=P/t$, $\mathbf{A}\in\mathbb{C}^{a\times b}$, $\mathbf{R}_a$ is an $a\times a$ matrix, $\lambda_k$ is the $k$th largest eigenvalue of $\mathbf{A}\mathbf{A}^{\dagger}$ (or $\mathbf{A}^{\dagger}\mathbf{A}$), $n_1=\min(s_2,e)$, and $f_{\lambda_k}(x)$ is given in Theorem 3. Note that the ergodic secrecy rate function takes an integral form rather than a closed form because $f_{\lambda_k}(x)$ is very complicated.

\vspace{0.1in}
\textsl{Theorem 3}: For $k=1,...,n$, the marginal pdf of the $k$th largest eigenvalue $\lambda_k$ of a receiver-side correlated central Wishart matrix $\mathbf{W}\sim W_n(m,\mathbf{0}_n,\mathbf{R}_a)$ is given by
\begin{equation}
f_{\lambda_k}(x)=K^{-1}\sum_{i=1}^k\sum_{\bm{\mu}\in\mathcal{P}(i)}\sum_{j=1}^{n}\det\big[\mathbf{G},\mathbf{\Omega}(\bm{\mu},\bm{\sigma},i,j;x)\big],	
\end{equation}
where we have
\begin{equation}\label{k}
K=\prod_{i<j}^n\sigma_i-\sigma_j \prod_{i=1}^n(b-i)!,
\end{equation}
and $\mathcal{P}(i)$ is a set of all permutations $(\mu_1,...,\mu_n)$ of integers $(1,...,n)$ such that $(\mu_1<\mu_2<...<\mu_{i-1})$ and $(\mu_i<\mu_{i+1}<...<\mu_n)$. The set has $\binom{n}{i-1}$ permutations of $\bm{\mu}$, each of which is a representation of the matrix function $\mathbf{\Omega}(\cdot)$. Hence, $\sum_{\bm{\mu}\in\mathcal{P}(i)}$ denotes a summation over these $\binom{n}{i-1}$ matrices. $\mathbf{G}$ is an $a\times(a-n)$ matrix, whose $(i,j)$th element is $\sigma_i^{j-1}$. Note that $\mathbf{G}$ is a null matrix when $b\geq a$. The $a \times n$ real matrix $\mathbf{\Omega}(\bm{\mu},\bm{\sigma},i,j;x)$ is defined as
\begin{flalign}\label{Matrixim}
&\big[\mathbf{\Omega}(\bm{\mu},\bm{\sigma},i,j;x)\big]_{u,\mu_v}\\ \notag
&\!\!\!=\begin{cases}
\sigma_u^{a-n+\mu_v-1}\Gamma(b-n+\mu_v,\frac{x}{\sigma_u}),&\!\!\!\text{$v=1,...,k-1$, $\mu_v \neq j$},\\
-\sigma_u^{a-b-1}\exp(-\frac{x}{\sigma_u})x^{b-n+\mu_v-1},&\!\!\!\text{$v=1,...,k-1$, $\mu_v = j$},\\
\sigma_u^{a-n+\mu_v-1}\gamma(b-n+\mu_v,\frac{x}{\sigma_u}),&\!\!\!\text{$v=k,...,n$, $\mu_v \neq j$},\\
\sigma_u^{a-b-1}\exp(-\frac{x}{\sigma_u})x^{b-n+\mu_v-1},&\!\!\!\text{$v=k,...,n$, $\mu_v = j$},
\end{cases}
\end{flalign}
for $u=1,...,a$ and $v=1,...,n$, where $\Gamma(\cdot,\cdot)$ and $\gamma(\cdot,\cdot)$ are the upper and lower incomplete Gamma functions \cite{horn2012matrix} defined as
\begin{flalign}
&\Gamma(\epsilon,x)=\int_x^{\infty}\exp(-z)z^{\epsilon-1}\text{d}z,\label{UGmma}\\
&\gamma(\epsilon,x)=\int_0^x\exp(-z)z^{\epsilon-1}\text{d}z.\label{LGmma}
\end{flalign}

\vspace{0.1in}
\textsl{Proof:} See Appendix. C.

\vspace{0.1in}
Note that when $\mathbf{R}_r=\mathbf{I}_r$ or $\mathbf{R}_e=\mathbf{I}_e$, $C_\mathbf{A}(\mathbf{I}_a,\rho,\eta)$ will be replaced by the equation in \cite[Eq. (17)]{Liuwishart2017}.

\vspace{0.1in}
\textsl{Remark 2}: We can use Eqn. (\ref{esc}), as a theoretical ergodic secrecy rate expression of Eqn. (\ref{lowesc}), to maximize the ergodic secrecy rate via a one-dimensional search, which takes the number of eigen-subchannels of message streams, i.e., $s_1$, as a search direction. Although the results from the search are not globally optimal and the achieved ergodic secrecy rates are the lower bounds of ergodic secrecy capacities, the search with its complexity $O(n)$ avoids complicated convex optimization processes.

\vspace{0.1in}
The eigen-subchannels of larger eigenvalues should be selected for sending messages because $C_{\mathbf{H}}(\mathbf{R}_r, \rho,s_1)$ in Eqn. (\ref{esc}) is larger when using the eigen-subchannels of larger eigenvalues for a fixed $s_1$. Meanwhile, which one is selected for AN signals has no effect on the ergodic secrecy rate for a fixed $s_1$, because $C_{\mathbf{H}_3}(\mathbf{R}_e, \rho,n_1)$ in Eqn. (\ref{esc}) is a constant for a given $s_1$. In addition, $C_{\mathbf{H}_4}(\mathbf{R}_e,\rho,e)$ in Eqn. (\ref{esc}) is an average value over $\mathbf{H}_4$, and is fixed for given $t$ and $e$, which have nothing to do with $s_1$. Hence, the optimal method must be that the eigen-subchannels for messages are selected from their large to small corresponding eigenvalues. For example, given $t=4$, $s_1=2$, and $s_2=2$, the maximization is achieved if the 1st and 2nd eigen-subchannels are selected for sending messages, while the 3rd and 4th eigen-subchannels are selected for sending AN signals. In this case, maximization is done over an array with $n$ elements, and the eigen-subchannel allocation is a one-dimensional search problem with its complexity $O(n)$, where $n=\min(t,r)$.

\subsection{Approximate Ergodic Secrecy Rate}
The derived ergodic secrecy rate expression in Eqn. (\ref{esc}) is not in a closed form. We can simplify the expression to an approximate form, to show the impacts of correlated matrices $\mathbf{R}_r$ (a function of $d_{\text{Bob}}, \bar{\theta}_{\text{Bob}}, \text{and } \delta_{\text{Bob}}$) and $\mathbf{R}_e$ (a function of $d_{\text{Eve}}, \bar{\theta}_{\text{Eve}}, \text{and } \delta_{\text{Eve}}$) on the ergodic secrecy rates. 

\vspace{0.1in}
\textsl{Theorem 4:} The ergodic secrecy rate, i.e., Eqn. (\ref{esc}), can be expressed approximately as
\begin{flalign}
R^{\text{app}}_s&=[\chi_1+\chi_2]^{+}, \label{R43}
\end{flalign}
where 
\begin{flalign}\label{chi1}
&\chi_1=\sum_{i=1}^{s_1}\log_2\big\{1+\rho\text{E}[\lambda_i(\mathbf{H}\mathbf{H}^{\dagger})]\big\},  \\
&\chi_2=\log_2\bigg[\frac{1+\sum_{k=1}^{e}\rho^k\prod_{i=0}^{k-1}(m_1-i)\varrho_k}{1+\sum_{k=1}^{e}\rho^k\prod_{i=0}^{k-1}(t-i)\varrho_k} \bigg], \label{chi2}\\
&\varrho_k=\sum_{\ell_1<\ell_2<...<\ell_k}\det[\mathbf{R}_{e,(\ell_1,...\ell_k)}], \label{varrho}\\
&\mathbf{R}_{e,(\ell_1,...\ell_k)}=
\left[
\begin{matrix} \label{reell}
[\mathbf{R}_e]_{(\ell_1,\ell_1)}&\ldots&[\mathbf{R}_e]_{(\ell_1,\ell_k)}\\
[\mathbf{R}_e]_{(\ell_2,\ell_1)}&\ldots&[\mathbf{R}_e]_{(\ell_2,\ell_k)}\\ 
\vdots& &\vdots\\
[\mathbf{R}_e]_{(\ell_k,\ell_1)}&\ldots&[\mathbf{R}_e]_{(\ell_k,\ell_k)}
\end{matrix}
\right],  \\
&n_1 =\min(e,d), \quad m_1=\max(e,d), \quad 1\leq k \leq e. \label{n1m1}
\end{flalign}
Here, we denote a subset $(\ell_1,...,\ell_k)$ of $(1,2,...,e)$ such that $\ell_1<\ell_2<...<\ell_k$, which means $\mathbf{R}_{e,(\ell_1,...\ell_n)}=\mathbf{R}_e$. Similar matrix structures and more explanations were given in \cite{Zhang2005} and \cite{Cui2005}. $\lambda_1(\mathbf{H}\mathbf{H}^{\dagger})>\lambda_2(\mathbf{H}\mathbf{H}^{\dagger})>...>\lambda_n(\mathbf{H}\mathbf{H}^{\dagger})$ are the ordered eigenvalues of $\mathbf{H}\mathbf{H}^{\dagger}$. 

\vspace{0.1in}
\textsl{Proof:} See Appendix. D.

\vspace{0.1in}
\textsl{Remark 3}: (The impact of $d_{\text{Bob}}, \bar{\theta}_{\text{Bob}}, \text{and} \delta_{\text{Bob}}$): $\mathbf{R}_r$ only affects $\chi_1$. When $s_1=1$, Eqn. (\ref{chi1}) can be re-written as
\begin{flalign}\label{chi11}
\chi_1=&\log_2\big\{1+P\text{E}[\lambda_1(\mathbf{H}\mathbf{H}^{\dagger}/t)]\big\}.
\end{flalign}
From \cite[Th. 1.1]{baik2006eigenvalues}, when $r/t=c<1$, $t\rightarrow +\infty$, and $r \rightarrow +\infty$, we can get
\begin{flalign}
\text{E}[\lambda_1(\mathbf{H}\mathbf{H}^{\dagger}/t)]&\rightarrow \begin{cases}
\sigma_1(1+\frac{c}{\sigma_1-1}),&\!\!\!\text{$\sigma_1>1+\sqrt{c}$},\\
(1+\sqrt{c})^2,&\!\!\!\text{$\sigma_1\leq 1+\sqrt{c}$},
\end{cases}
\end{flalign}
where $\sigma_1$ is the largest eigenvalue of $\mathbf{R}_r$, i.e., $\lambda_1(\mathbf{R}_r)$. Based on Theorem 1, we know that $\sigma_1$ decreases monotonically with increasing $d_{\text{Bob}}$, $\bar{\theta}_{\text{Bob}}$, and $\delta_{\text{Bob}}$, respectively. Thus, $\text{E}[\lambda_1(\mathbf{H}\mathbf{H}^{\dagger}/t)]$ decreases monotonically with increasing $d_{\text{Bob}}$, $\bar{\theta}_{\text{Bob}}$, and $\delta_{\text{Bob}}$, respectively, and then keeps constant. We can conclude that, when $s_1=1$, an ergodic secrecy rate decreases monotonically with increasing $d_{\text{Bob}}$, $\bar{\theta}_{\text{Bob}}$, and $\delta_{\text{Bob}}$, respectively.

When $s_1=n=\min(t,r)$, based on \cite[Eqs. (22) and (30)]{Zhang2005}, Eqn. (\ref{chi1}) can be expressed approximately as
\begin{flalign}\label{chi12}
\chi_1&=\sum_{i=1}^n \log_2\big\{1+\rho\text{E}[\lambda_i(\mathbf{H}\mathbf{H}^{\dagger})]\big\}, \\ \notag
&\simeq n\log_2\rho+\hbar+\log_2\det[\mathbf{R}_r],
\end{flalign}
where
\begin{flalign}
\hbar=\begin{cases}
\log_2\sum_{i=0}^{n-1}(m-i)&\text{$\rho \text{E}[\lambda_i(\mathbf{H}\mathbf{H}^{\dagger})]\gg1$},\\
\sum_{i=0}^{n-1}\psi(m-i)&\text{$\rho \text{E}[\lambda_i(\mathbf{H}\mathbf{H}^{\dagger})]\ll1$},
\end{cases}
\end{flalign}
and $m=\max(t,r)$, $n=\min(t,r)$, and $\psi(x)$ is defined as
\begin{equation}
\psi(x)=-\xi+\sum_{i=1}^{x-1}\frac{1}{i},
\end{equation}
where $\xi \simeq 0.5772156649$ is the Euler's constant. It is obvious that Eqn. (\ref{chi12}) increases monotonically with an increasing $\det[\mathbf{R}_r]$. Based on Theorem 1, we know that $\det[\mathbf{R}_r]$ increases monotonically with increasing $d_{\text{Bob}}$, $\bar{\theta}_{\text{Bob}}$, and $\delta_{\text{Bob}}$, respectively. In conclusion, when $s_1=n$, an ergodic secrecy rate increases monotonically with increasing $d_{\text{Bob}}$, $\bar{\theta}_{\text{Bob}}$, and $\delta_{\text{Bob}}$, respectively. However, when $n>s_1>1$, it is very hard to find a simple relationship between an ergodic secrecy rate and correlation parameters, and thus we simulate these scenarios, as given in Section IV.

\vspace{0.1in}
\textsl{Remark 4}: (The impact of $d_{\text{Eve}}, \bar{\theta}_{\text{Eve}}, \text{and } \delta_{\text{Eve}}$): In Eqn. (\ref{R43}), $\mathbf{R}_e$ affects $\chi_2$ only. Based on \cite [Eqs. (12) and (16)] {Zhang2005}, we can get
\begin{equation}
\det[\mathbf{I}+\mathbf{R}_e]=1+\sum_{k=1}^e\sum_{\ell_1<\ell_2<...<\ell_k}\det[\mathbf{R}_{e,(\ell_1,...\ell_k)}].
\end{equation}
Since $\mathbf{R}_e$ is a Hermitian positive definite matrix, $\det[\mathbf{I}+\mathbf{R}_e]$ increases monotonically with an increasing $\det[\mathbf{R}_e]$. Certainly, we know that $\sum_{k=1}^e\sum_{\ell_1<\ell_2<...<\ell_k}\det[\mathbf{R}_{e,(\ell_1,...\ell_k)}]$ increases monotonically with an increasing $\det[\mathbf{R}_e]$. In addition, we will introduce an auxiliary function $f(\mathbf{x})$ as
\begin{equation}
f(\mathbf{x})=\frac{1+\sum_{k=1}^e a_k x_k}{1+\sum_{k=1}^e b_k x_k},
\end{equation} 
where $b_k>a_k$, $\forall k$. $f(\mathbf{x})$ decreases monotonically with $\sum_{k=1}^e x_k$. Hence, $\chi_2$ decreases monotonically with $\sum_{k=1}^e\sum_{\ell_1<\ell_2<...<\ell_k}\det[\mathbf{R}_{e,(\ell_1,...\ell_k)}]$ as well as $\det[\mathbf{R}_e]$. Therefore, Eqn. (\ref{chi2}) decreases with $\det[\mathbf{R}_e]$. Similarly, $\det[\mathbf{R}_e]$ increases monotonically with $d_{\text{Eve}}$, $\bar{\theta}_{\text{Eve}}$, and $\delta_{\text{Eve}}$, respectively. Thus, an ergodic secrecy rate decreases monotonically with increasing $d_{\text{Eve}}$, $\bar{\theta}_{\text{Eve}}$, and $\delta_{\text{Eve}}$, respectively.

Table \ref{Cons} shows the impacts of $\{d_{\text{Bob}}, \bar{\theta}_{\text{Bob}}, \text{and } \delta_{\text{Bob}}\}$, as well as  $\{d_{\text{Eve}}, \bar{\theta}_{\text{Eve}}, \text{and } \delta_{\text{Eve}}\}$ on the ergodic secrecy rates. We use $\uparrow$ and $\downarrow$ to represent monotonically ``increase" and ``decrease", respectively. For example, ``$R^{\text{app}}_s\downarrow$ with $\{d_{\text{Eve}}, \bar{\theta}_{\text{Eve}}, \delta_{\text{Eve}}\}\uparrow$" means that ``the ergodic secrecy rate decreases monotonically with increasing $d_{\text{Eve}}$, $\bar{\theta}_{\text{Eve}}$, and $\delta_{\text{Eve}}$, respectively". We must point out that ``increase" or ``decrease" will not take place forever, because when the correlation parameters grow to a certain extent, the correlation disappears and ergodic secrecy rates will be constant. 

\begin{table}[t]
\renewcommand{\arraystretch}{1.1}
\newcommand{\PreserveBackslash}[1]{\let\temp=\\#1\let\\=\temp}
\caption{Impacts of correlation on ergodic secrecy rates.}\label{Cons}\centering
\begin{tabular}{C{2.3cm}@{}C{1.5cm}@{}C{4cm}}\toprule
 Sides & $s_1$ & Impacts\\  \midrule
\multirow{2}{*}{Correlation at Bob} & $s_1=1$ & $R_s^{\text{app}}\downarrow$ with $\{d_{\text{Bob}}, \bar{\theta}_{\text{Bob}}, \delta_{\text{Bob}}\}\uparrow$\\  
 & $s_1=n$ & $R_s^{\text{app}}\uparrow$ with $\{d_{\text{Bob}}, \bar{\theta}_{\text{Bob}}, \delta_{\text{Bob}}\}\uparrow$\\ 
Correlation at Eve & Arbitrary $s_1$ & $R^{\text{app}}_s\downarrow$ with $\{d_{\text{Eve}}, \bar{\theta}_{\text{Eve}}, \delta_{\text{Eve}}\} \uparrow$ \\
\bottomrule
\end{tabular}
\end{table}

\begin{figure*}[h]
\centering
\subfigure[Ergodic secrecy rates in low SNR regions.]{
\label{simpower1} 
\includegraphics[width=0.46\textwidth]{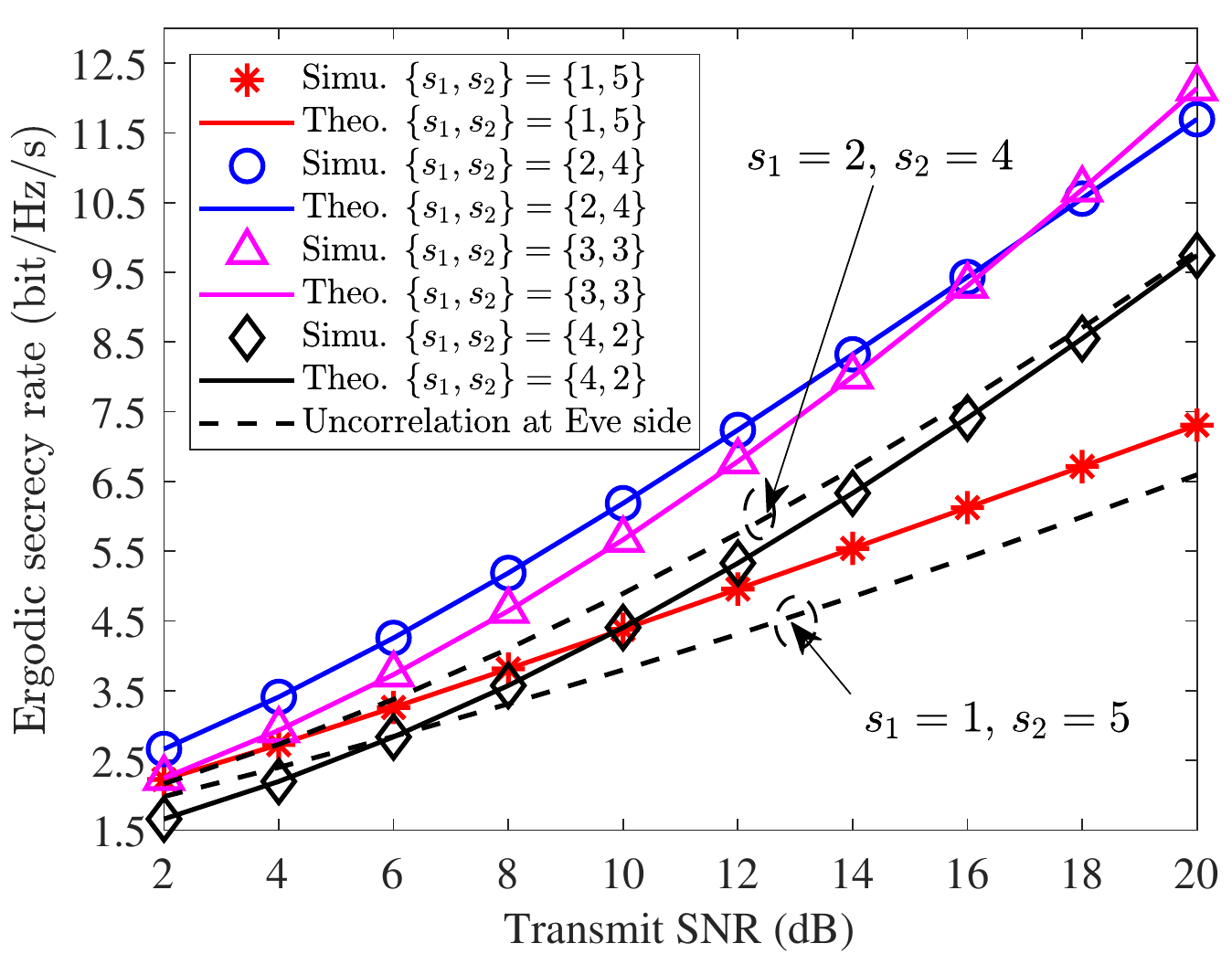}}
\hspace{1.1cm}
\subfigure[Ergodic secrecy rates in high SNR regions.]{
\label{simpower2} 
\includegraphics[width=0.45\textwidth]{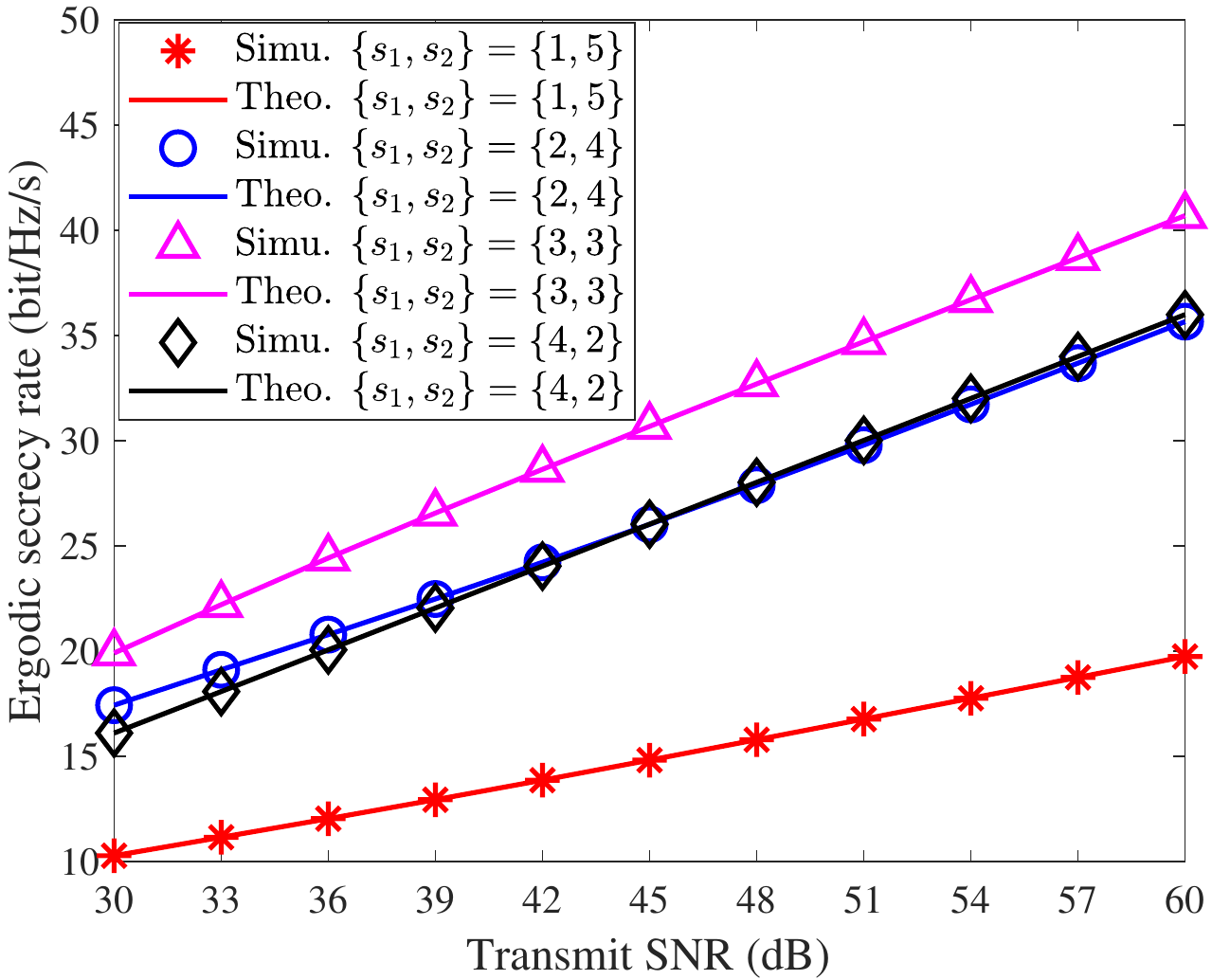}}
\caption{Numerical and simulation results of ergodic secrecy rates of a correlated MIMO channel in terms of transmit SNR, where $t=6$, $r=e=4$, $d_{\text{Bob}}=d_{\text{Eve}}=0.8$, $\bar{\theta}_{\text{Bob}}=\bar{\theta}_{\text{Eve}}=30^{\circ}$, and $\delta_{\text{Bob}}=\delta_{\text{Eve}}=10^{\circ}$.}
\label{simpower} 
\end{figure*}

\section{NUMERICAL AND SIMULATE RESULTS}	

In this section, numerical and simulation results are given. As shown in the figures below, the theoretical results (theo.) from Eqn. (\ref{esc}) are in a good agreement with the Monte Carlo simulations (simu.) of $10^5$ independent runs on Eqn. (\ref{ISC1}). The ergodic secrecy rates of the proposed scheme are compared to the traditional AN schemes \cite{Tsai2014, Liu2015, Yun2017conf,Yun2018}, which did not consider the correlation and used all eigen-subchannels to transmit messages, i.e., $s_1=n$. In the proposed scheme, the number of eigen-subchannels for sending messages, i.e., $s_1$, is a variable. The channel model in the simulations is a receiver-side correlated Rayleigh fading channel.

Fig. \ref{simpower} illustrates the impact of transmit SNR on ergodic secrecy rates with different choices of $\{s_1$, $s_2\}$. As shown in Fig. \ref{simpower1}, the achievable ergodic secrecy rates increase almost exponentially with SNR, and $s_1=2$ is the best choice when SNR$<16$. There exists a crossing point between $s_1=2$ and $s_1=3$ because $s_1=3$ offers a better performance with an increasing SNR, which is consistent with \cite [Th. 5] {Liuwishart2017}. In addition, the black and dashed lines are simulation results without the awareness of the correlated fading that are conformed to the scenarios $\{s_1=2,s_2=4\}$ and $\{s_1=1,s_2=5\}$\footnote{The more results are given in \url{https://github.com/yiliangliu1990/liugit_pub}.}. If we do not consider (or do not know) correlation parameters at Eve's sides, the ergodic secrecy rates will be reduced compared to the performance with the knowledge of Eve's correlation parameters, because the ergodic wiretap channel rate will be enlarged if there is on correlation among Eve's antennas.

Fig. \ref{simpower2} shows the results in high SNR regions, where ergodic secrecy rates grow almost linearly with SNR. We see that $s_1=3$ is the best choice, and the simulations of $s_1=2$ and $s_1=4$ show similar performance. The results indicate that it is better to allocate stronger eigen-subchannels to transmit messages and weaker eigen-subchannels to send AN signals, especially in high SNR regions, which coincides with the results given in the uncorrelated scenarios \cite{Liuwishart2017}. 

\begin{figure}[h]
\centering
\label{simant1} 
\includegraphics[width=0.5\textwidth]{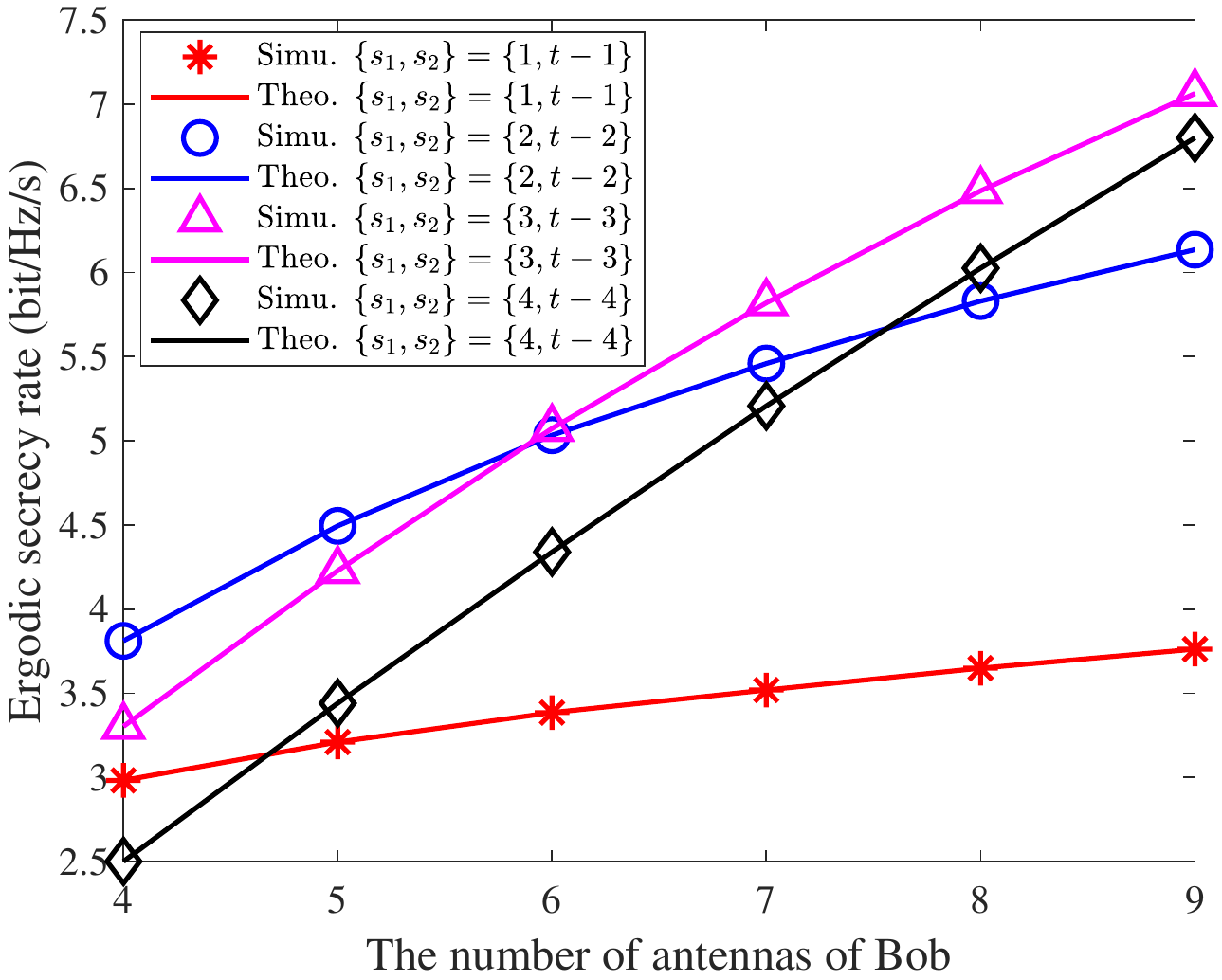}
\caption{Numerical and simulation results of ergodic secrecy rates in a correlated MIMO channel in terms of the number of antennas of Bob, where $t=5$, $e=3$, transmit SNR=5 dB, $d_{\text{Bob}}=d_{\text{Eve}}=0.8$, $\bar{\theta}_{\text{Bob}}=\bar{\theta}_{\text{Eve}}=30^{\circ}$, and $\delta_{\text{Bob}}=\delta_{\text{Eve}}=10^{\circ}$.}
\label{simant} 
\end{figure}

Fig. \ref{simant} shows the relationship between ergodic secrecy rates and the number of antennas of Bob. We can observe that an increasing number of antennas at Bob is beneficial for any $s_1$ and $s_2$ chosen. If Bob has many receive antennas, it can enlarge the channel gains of the message streams because Bob has enough antennas to decode them and gather the received power from all antennas. The previous works in \cite{Tsai2014, Liu2015, Yun2017conf,Yun2018} did not consider the scenarios with $t<r$, and thus we do not compare them here.

\begin{figure*}[t]
\centering
\subfigure[Ergodic secrecy rates in terms of $d_{\text{Bob}}$, where $d_{\text{Eve}}=0.8$. ]{
\label{simdis1} 
\includegraphics[width=0.45\textwidth]{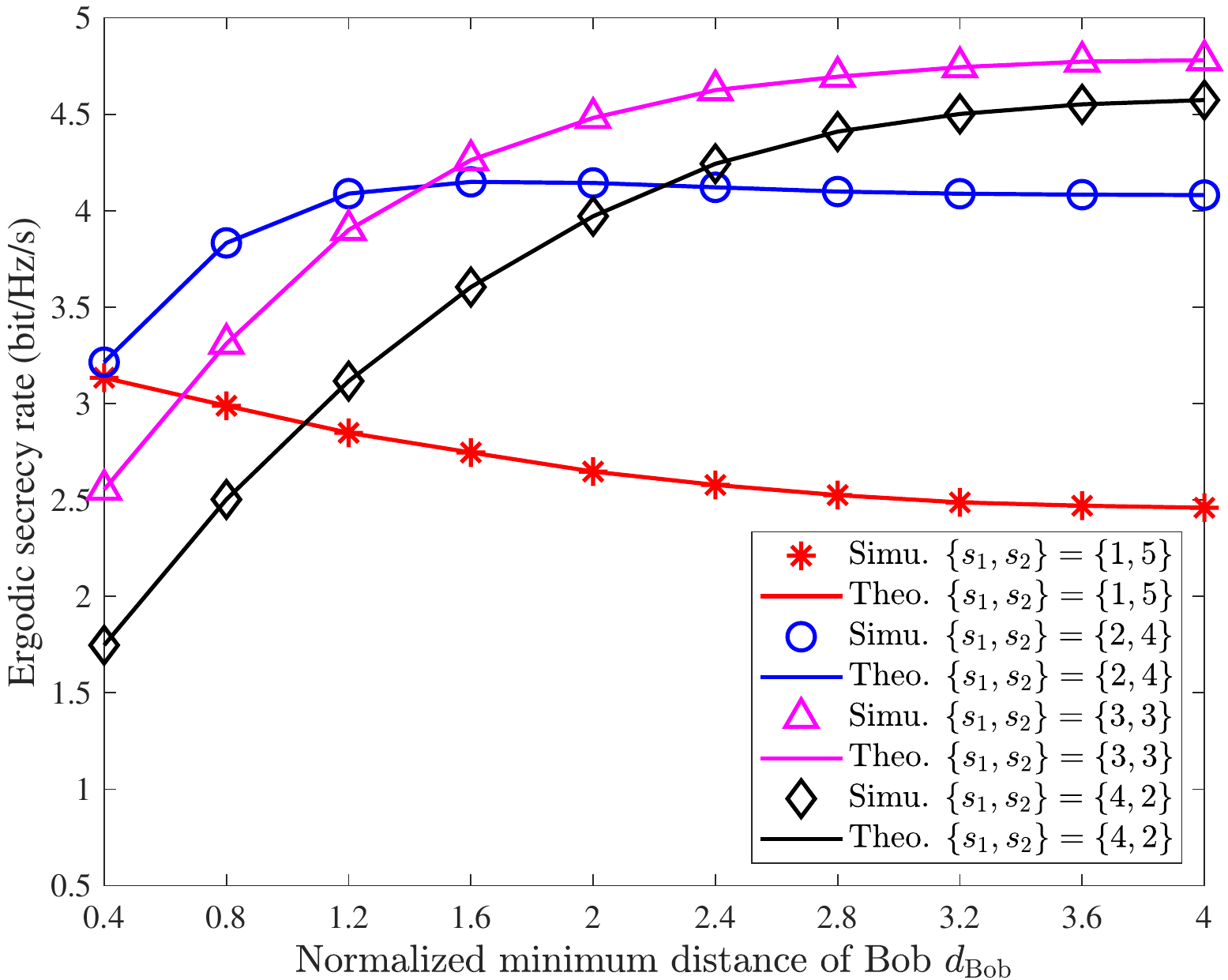}}
\hspace{1.1cm}
\subfigure[Ergodic secrecy rates in terms of $d_{\text{Eve}}$, where $d_{\text{Bob}}=0.8$.]{
\label{simdis2} 
\includegraphics[width=0.45\textwidth]{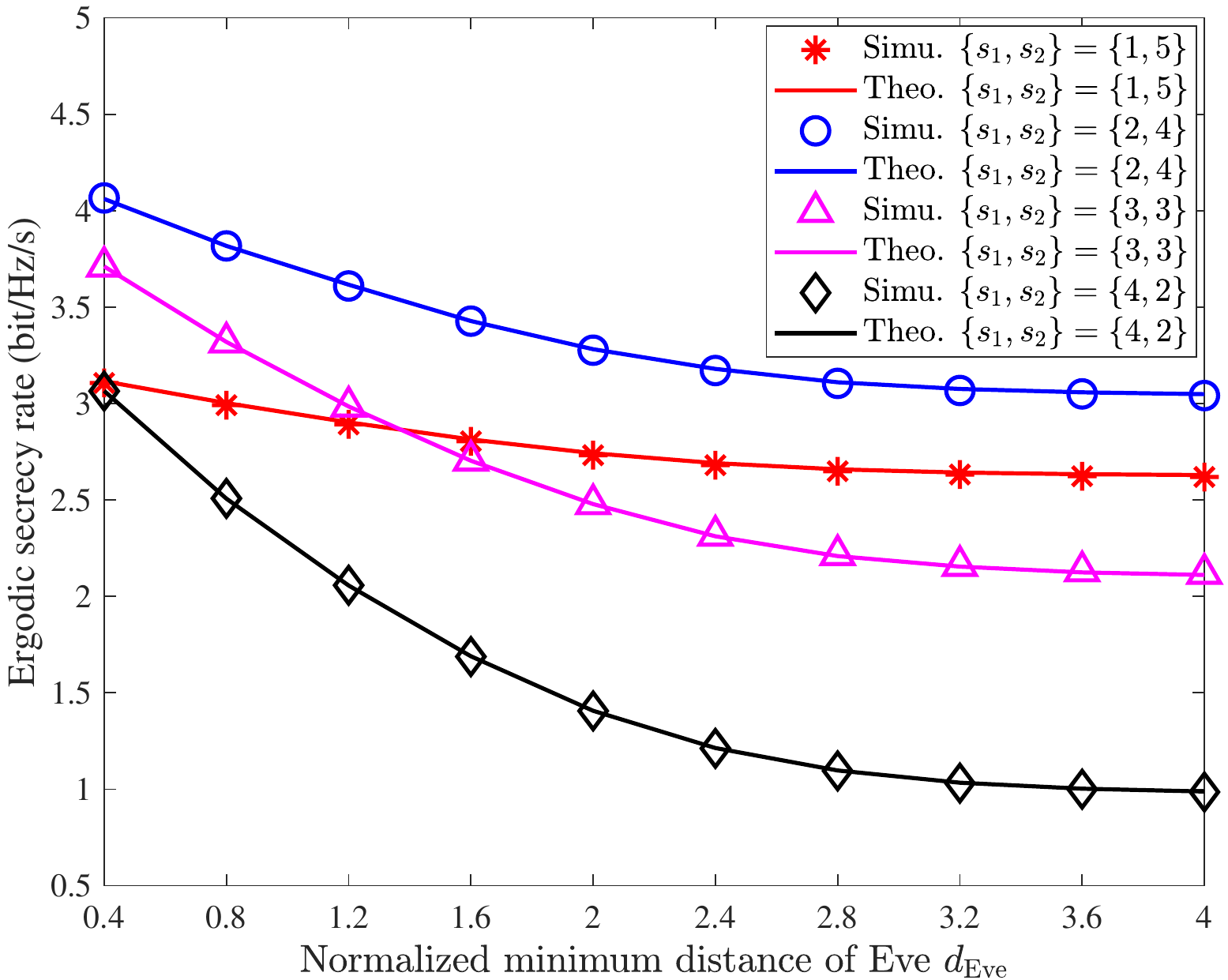}}
\caption{Numerical and simulation results of ergodic secrecy rates in a correlated MIMO channel in terms of the normalized minimum distances $d_{\text{Bob}}$ and $d_{\text{Eve}}$, where transmit SNR=5 dB, $t=6$, $r=e=4$, $\bar{\theta}_{\text{Bob}}=\bar{\theta}_{\text{Eve}}=30^{\circ}$, and $\delta_{\text{Bob}}=\delta_{\text{Eve}}=10^{\circ}$.}
\label{simdis} 
\end{figure*}

\begin{figure*}
\centering
\subfigure[Ergodic secrecy rates in terms of $\bar{\theta}_{\text{Bob}}$, where $\bar{\theta}_{\text{Eve}}=30^{\circ}$.]{
\label{simaoa1} 
\includegraphics[width=0.45\textwidth]{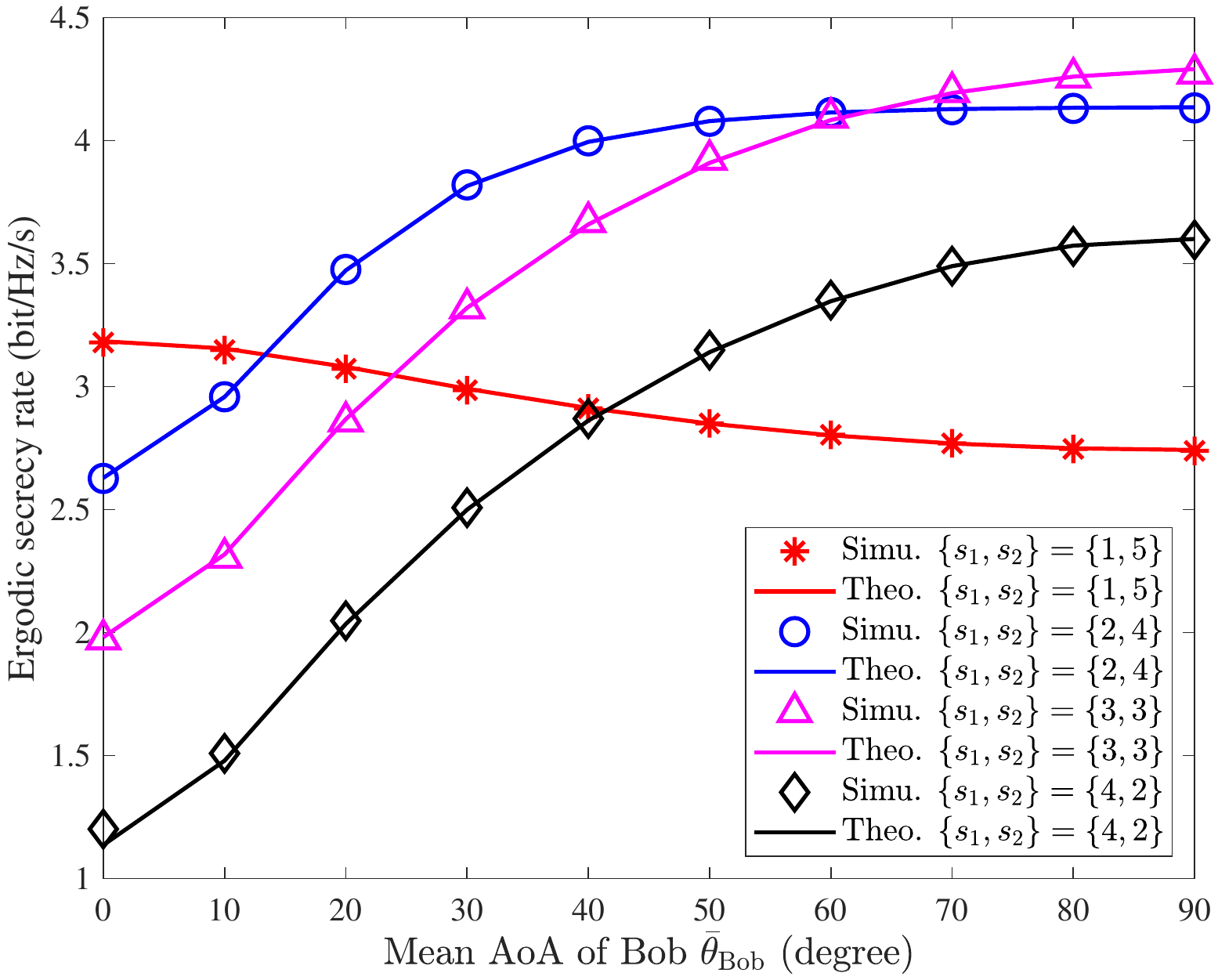}}
\hspace{1.1cm}
\subfigure[Ergodic secrecy rates in terms of $\bar{\theta}_{\text{Eve}}$, where $\bar{\theta}_{\text{Bob}}=30^{\circ}$.]{
\label{simaoa2} 
\includegraphics[width=0.45\textwidth]{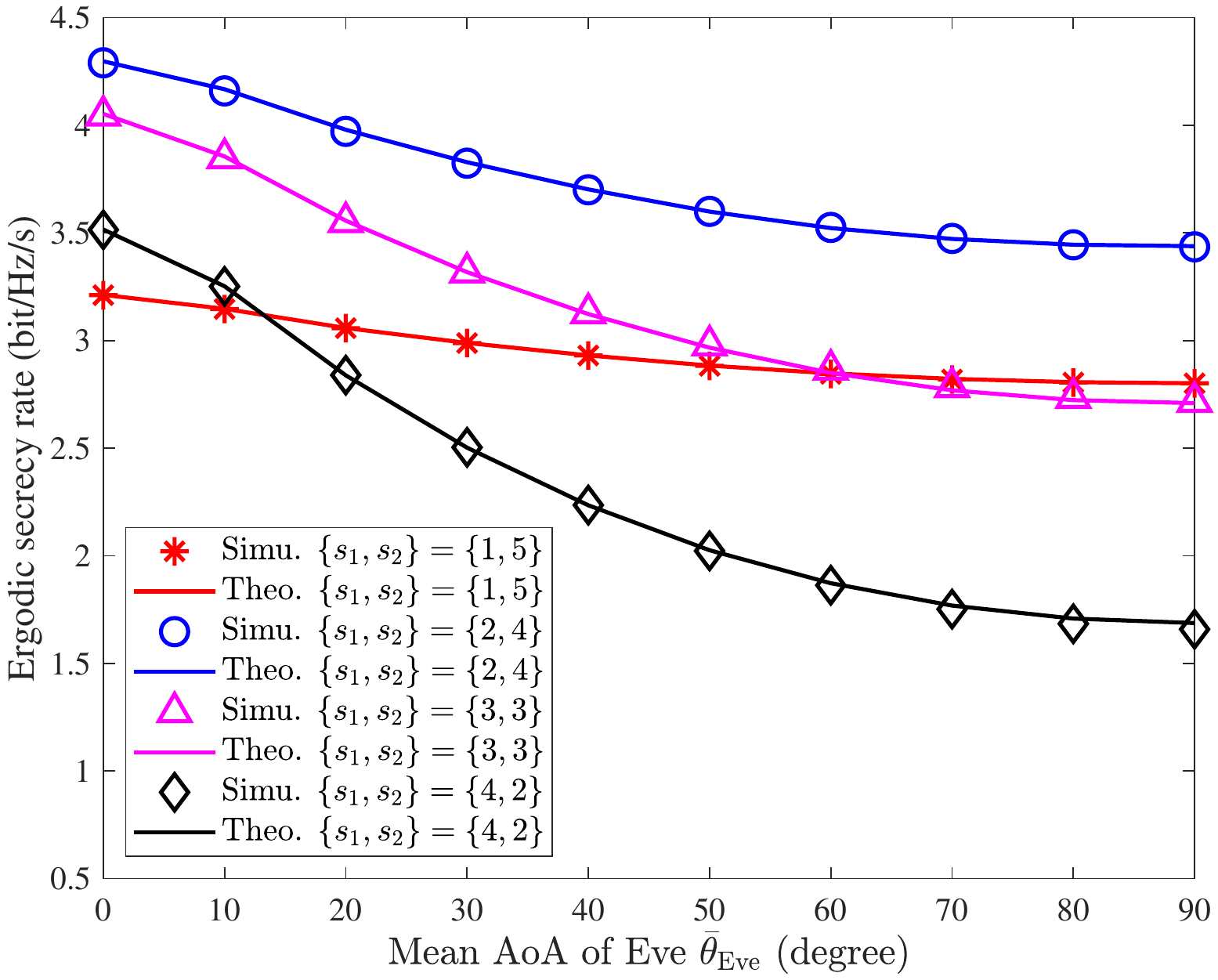}}
\caption{Numerical and simulation results of ergodic secrecy rates in a correlated MIMO channel in terms of mean AoA $\bar{\theta}_{\text{Bob}}$ and $\bar{\theta}_{\text{Eve}}$, where transmit SNR$=$5 dB, $t=6$, $r=e=4$, $d_{\text{Bob}}=d_{\text{Eve}}=0.8$, and $\delta_{\text{Bob}}=\delta_{\text{Eve}}=10^{\circ}$.}
\label{simaoa} 
\end{figure*}

\begin{figure*}
\centering
\subfigure[Ergodic secrecy rates in terms of $\delta_{\text{Bob}}$, where $\delta_{\text{Eve}}=10^{\circ}$. ]{
\label{simras1} 
\includegraphics[width=0.45\textwidth]{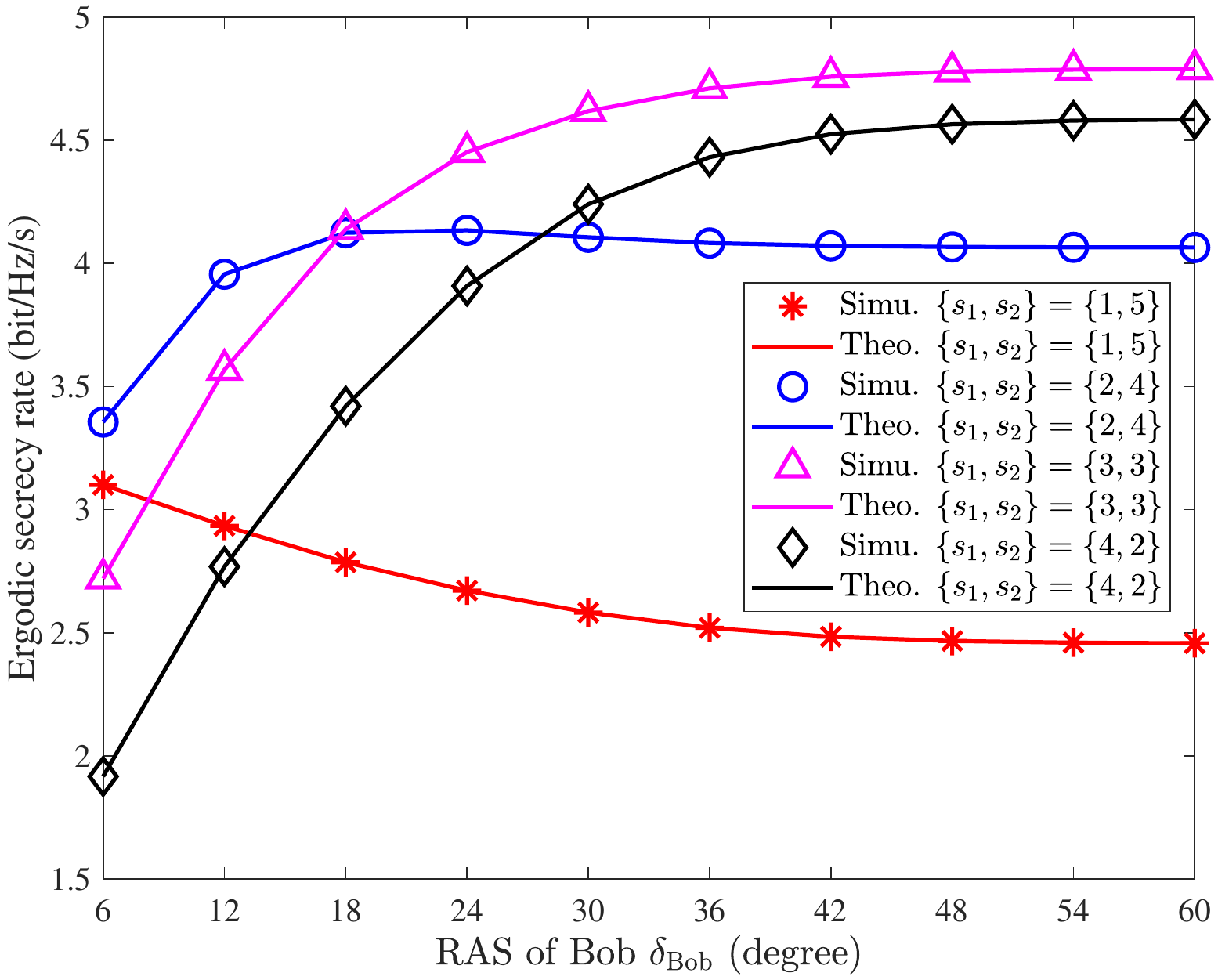}}
\hspace{1.1cm}
\subfigure[Ergodic secrecy rates in terms of $\delta_{\text{Eve}}$, where $\delta_{\text{Bob}}=10^{\circ}$.]{
\label{simras2} 
\includegraphics[width=0.45\textwidth]{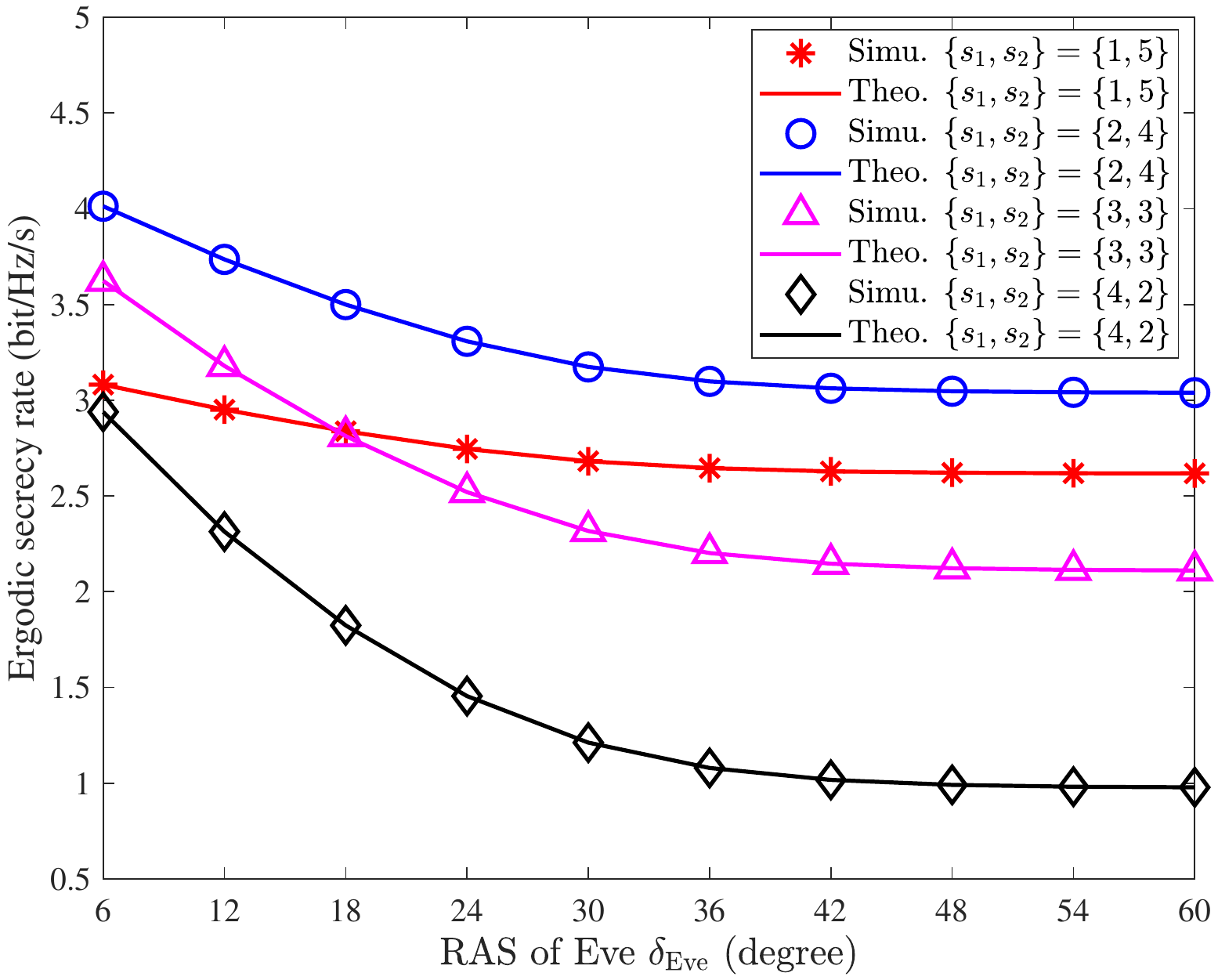}}
\caption{Numerical and simulation results of ergodic secrecy rates in a correlated MIMO channel in terms of RAS $\delta_{\text{Bob}}$ and $\delta_{\text{Eve}}$, where transmit SNR$=$5 dB, $t=6$, $r=e=4$, $d_{\text{Bob}}=d_{\text{Eve}}=0.8$, and $\bar{\theta}_r=\bar{\theta}_e=30^{\circ}$.}
\label{simras} 
\end{figure*}

Fig. \ref{simdis} shows the ergodic secrecy rate simulations in terms of antenna spacing in wavelength, where we set SNR$=5$ dB, $\bar{\theta}_{\text{Bob}}=\bar{\theta}_{\text{Eve}}=30^{\circ}$, and $\delta_{\text{Bob}}=\delta_{\text{Eve}}=10^{\circ}$. Assume that $d_{\text{Eve}}$ is fixed in Fig. \ref{simdis1}. When $s_1=3$ and $s_1=4$, the ergodic secrecy rates grow with the antenna spacing; when $s_1=1$, the ergodic secrecy rates decrease with the antenna spacing. If $s_1=2$, we see a peak value of the ergodic secrecy rates, where the rates rise at the beginning, then reduce, and keep constant. Intuitively, the peak occurs due to the fact that the curve of $s_1=2$ is affected by variations of the largest and the second largest eigen-subchannels, where the gain of the second largest eigen-subchannel increases fast at the beginning, and the gain of the largest eigen-subchannel decreases fast in the second half of the simulation diagram. In Fig. \ref{simdis2}, we assumed that $d_{\text{Bob}}$ is fixed, and we can see that the ergodic secrecy rates decrease with the antenna spacing. In particular, the more eigen-subchannels are allocated for messages, the more quickly the ergodic secrecy rates will decrease. Note that in Figs. \ref{simdis1} and \ref{simdis2}, when the normalized minimum distances are larger than three, the receiver-side correlation almost disappears. Therefore, the curves of ergodic secrecy rates tend to be constant. This phenomenon also appears in traditional MIMO systems\cite{Simon2006,Ghaderipoor2012}.

Fig. \ref{simaoa} examines the ergodic secrecy rates in terms of mean AoA in a correlated MIMO channel, where $d_{\text{Bob}}=d_{\text{Eve}}=0.8$ and $\delta_{\text{Bob}}=\delta_{\text{Eve}}=10^{\circ}$. Fig. \ref{simaoa1} shows the effects of Bob's mean AoA with a fixed $\bar{\theta}_{\text{Eve}}$, where an increasing mean AoA will reduce the gain of the strongest eigen-subchannel, such that the ergodic secrecy rates will be reduced if we only chose the strongest one, i.e., $s_1=1$. However, an increasing mean AoA will reduce the correlation at receiver-side, and thus the ergodic secrecy rate will increase if we use most of the eigen-subchannels to transmit messages. As shown in Fig. \ref{simaoa2}, assuming $\bar{\theta}_{\text{Bob}}$ is fixed, we see that the ergodic secrecy rates will reduce with an increasing AoA of Eve because an increasing AoA of Eve will reduce the receiver-side correlation, which enlarges the wiretap channel capacities but does not affect the main channel capacities at all.

RAS has also its impact on the ergodic secrecy rates, which has a similar effect as the mean AoA. As shown in Fig. \ref{simras1}, an increasing RAS of Bob will reduce receiver-side correlation and reduce the gain of the strongest eigen-subchannel. Hence, when $s_1=1$, the ergodic secrecy rates will decrease with $\delta_{\text{Bob}}$, and when $s_1=2, 3, \text{and }4$, the ergodic secrecy rates will increase because a weaker receiver-side correlation enlarges the main channel capacities. As shown in Fig. \ref{simras2} with a fixed $\delta_{\text{Bob}}$, an increasing RAS of Eve reduces the ergodic secrecy rate with an arbitrary number of message streams. The curve of $s_1=4$ grows fast if compared to $s_1$=1, 2, and 3. 

\section{CONCLUSIONS AND FUTURE WORKS}
In this paper, we investigated the ergodic secrecy rate of spatially correlated scattering Rayleigh fading channels in an artificial noisy MIMO system, along with theoretical and approximate ergodic secrecy rate analysis. The suitable number of eigen-subchannels for sending messages and AN signals can be identified via a one-dimensional search based on the derived ergodic secrecy rate expressions. According to the results given in the analyses and simulations, we revealed that the correlation parameters, i.e., mean AoA, RAS, and antenna spacing, have significant influence on ergodic secrecy rates. Nevertheless, a real MIMO channel may be transmitter-side correlated or doubly-correlated at both sides. When considering the transmitter-side correlated or doubly-correlated channels, the derivation of statistical distribution of $\mathbf{H}_e\mathbf{F}$, as shown in Theorem 2, is still an open issue. Hence, in the future, we should establish a new Wishart matrix model first before investigating the impacts of those correlated channels on ergodic secrecy rates.

\section{APPENDICES}
\subsection{Proof of Theorem 1}
\textsl{Lemma 1} (Proved in \cite [Th. 2.1] {HORN1992337}): For $r\times r$ matrices $\mathbf{A}=[a_{ij}]$ and $\mathbf{B}=[b_{ij}]$, if $\mathbf{A}$ and $\mathbf{B}$ are Hermitian positive (semi-)definite, then
\begin{flalign}
\sigma_1(\mathbf{A}\circ \mathbf{B})\leq \max_{1\leq i \leq r}a_{ii}\sigma_1(\mathbf{B}),
\end{flalign}
where $a_{ii}$ is a diagonal element of $\mathbf{A}$, and ``$\circ$" denotes the Schur product defined as $\mathbf{A} \circ \mathbf{B}=[a_{ij}b_{ij}]$.

\textsl{Lemma 2} (Proved in \cite [Th. 3] {BAPAT1985107}): For two $r\times r$ matrices $\mathbf{A}=[a_{ij}]$ and $\mathbf{B}=[b_{ij}]$, if $\mathbf{A}$ and $\mathbf{B}$ are Hermitian positive (semi-)definite, then
\begin{flalign}
\prod_{i=1}^r\sigma_i(\mathbf{A} \circ \mathbf{B}) \geq \prod_{i=1}^r\sigma_i(\mathbf{B})a_{ii}.
\end{flalign}

We begin to prove Theorem 1 as follows. For $d_1>d_2$, we can build up $\mathbf{R}_a(d_1)$ via $\mathbf{R}_a(d_2)$ as
\begin{equation}\label{r1r2tran}
\mathbf{R}_a(d_1)=\mathbf{M} \circ \mathbf{R}_a(d_2),
\end{equation}
where $\mathbf{M}$ is a Hermitian matrix whose diagonal elements are all one. If $d_1>d_2$ and $i \neq j$, based on Eqn. (\ref{corrematrix}), we can find that the modulus value of $[\mathbf{R}_a(d_1)]_{i,j}$ is smaller, i.e., 
\begin{equation}
|[\mathbf{R}_a(d_1)]_{i,j}|<|[\mathbf{R}_a(d_2)]_{i,j}|.
\end{equation}
Thus, $\mathbf{M}$ is positive (semi-)definite because all diagonal elements of $\mathbf{M}$ are one, and the modulus of non-diagonal elements is smaller than one. Based on Lemma 1 and Eqn. (\ref{r1r2tran}), we can get
\begin{flalign}
\sigma_1[\mathbf{R}_a(d_1)]\leq \max_{1\leq i \leq n}m_{ii}\sigma_1[\mathbf{R}_a(d_2)],
\end{flalign}
and $m_{ii}$ is a diagonal element of $\mathbf{M}$ such that $m_{ii}=1$. Hence, $\sigma_1[\mathbf{R}_a(d_1)]<\sigma_1[\mathbf{R}_a(d_2)]$. With the same argument, we can show that $\sigma_1[\mathbf{R}_a(\bar{\theta})]$ and $\sigma_1[\mathbf{R}_a(\delta)]$ have the same property. 

Based on Lemma 2 and $a_{ii}=1$, we get
\begin{flalign}
\det[\mathbf{R}_a(d_1)]&=\prod_{i=1}^r\sigma_i\big[\mathbf{M} \circ \mathbf{R}_a(d_2)\big] \\ \notag
&\geq\prod_{i=1}^r\sigma_i\big[\mathbf{R}_a(d_2)\big]m_{ii} =\det[\mathbf{R}_a(d_2)].
\end{flalign}
Note $\det[\mathbf{R}_a(d_1)]\neq \det[\mathbf{R}_a(d_2)]$, and thus ``$>$" is held. Similarly, $\det[\mathbf{R}_a(\bar{\theta})]$ and $\det[(\mathbf{R}_a(\delta)]$ have the same property. 

\hfill $\blacksquare$

\subsection{Proof of Theorem 2}
\textsl{Lemma 3} (Proved in \cite [Th. 2.3.2] {gupta1999matrix}): If $\mathbf{H}_e$$\sim$ $\mathcal{CN}_{e,t}(\mathbf{0},\mathbf{R}_e\otimes \mathbf{I}_t)$, the characteristic function of $\mathbf{H}_e$ is 
\begin{flalign}
\phi_{\mathbf{H}_e}(\mathbf{X})=\text{E}\big\{\text{etr}[i\mathbf{H}_e\mathbf{X}^{\dagger})]\big\}=\text{etr}\big(-\frac{1}{2}\mathbf{X}^{\dagger}\mathbf{R}_e\mathbf{X}\mathbf{I}_t\big),
\end{flalign}
where $i=\sqrt{-1}$.

Next, we can prove Theorem 2 based on Lemma 3. For a given $(t\times s)$ unitary matrix $\mathbf{B}$, the characteristic function of $\mathbf{H}_e\mathbf{B}$ is 
\begin{flalign}\label{changema}
\phi_{\mathbf{H}_e\mathbf{B}}(\mathbf{X})=\text{E}[\text{etr}(i\mathbf{H}_e\mathbf{B}\mathbf{X}^{\dagger})]=\text{E}[\text{etr}(i\mathbf{H}_e\mathbf{Y}^{\dagger})],
\end{flalign}
where $\mathbf{Y}^{\dagger}=\mathbf{B}\mathbf{X}^{\dagger}$. Viewing $\mathbf{Y}$ as a variable, from Lemma 1, we get
\begin{flalign}
\text{E}[\text{etr}(i\mathbf{H}_e\mathbf{Y}^{\dagger})]&=\text{etr}\big(-\frac{1}{2}\mathbf{Y}^{\dagger}\mathbf{R}_e\mathbf{Y}\big)\\ \notag
&=\text{etr}\big(-\frac{1}{2}\mathbf{X}^{\dagger}\mathbf{R}_e\mathbf{X}\mathbf{B}^{\dagger}\mathbf{B}\big).
\end{flalign}

Since $\mathbf{B}$ is a $(t\times s)$ unitary matrix, we have $\mathbf{B}^{\dagger}\mathbf{B}=\mathbf{I}_s$. Then, Eqn. (\ref{changema}) can be written as
\begin{flalign}\label{cf1}
\phi_{\mathbf{H}_e\mathbf{B}}(\mathbf{X})&=\text{etr}\big(-\frac{1}{2}\mathbf{X}^{\dagger}\mathbf{R}_e\mathbf{X}\mathbf{B}^{\dagger}\mathbf{B}\big)\\ \notag
&=\text{etr}\big(-\frac{1}{2}\mathbf{X}^{\dagger}\mathbf{R}_e\mathbf{X}\mathbf{I}_s\big).
\end{flalign}
As Eqn. (\ref{cf1}) is the characteristic function of a complex Gaussian matrix with its covariance matrix $\mathbf{R}_e\otimes\mathbf{I}_s$, the proof is completed. \hfill $\blacksquare$

\subsection{Proof of Theorem 3}
Let us define the cdf $F_{\lambda_k}(x)$ as
\begin{flalign}\label{proba0}
F_{\lambda_k}(x)&=P(\lambda_k\leq x)\\ \notag
&=P(\lambda_{k-1}\leq x)+p,
\end{flalign}
where $p=P(\lambda_n<\dots<\lambda_k<x<\lambda_{k-1}<\dots<\lambda_1)$. Let the domain be $D_1=\{0<\lambda_1<\dots<\lambda_n<x\}$, $D_2=\{x<\lambda_1<\dots<\lambda_n<\infty\}$, and $D_3=\{\lambda_n<\dots<\lambda_k<x<\lambda_{k-1}<\dots<\lambda_1\}$. 

\vspace{0.125in}
\textsl{Lemma 6} (Proved in \cite{James1964}): The joint pdf of the ordered eigenvalues $\lambda_1>\cdots>\lambda_n>0$ of a receiver-side correlated central Wishart matrix $\mathbf{W}\sim W_n(m,\mathbf{0}_n,\mathbf{R}_a)$ is
\begin{equation}\label{Base1}
f_{\bm{\lambda}}(\bm{\lambda})=K_0^{-1}\det\big[\mathbf{G}, \mathbf{E}(\bm{\lambda})\big]\prod_{i<j}^n(\lambda_i-\lambda_j)\prod_{i=1}^n\lambda_i^{b-n},
\end{equation}
where
\begin{flalign}
K_0=\begin{cases}
\prod_{i=1}^a \sigma_i^{b-n}(b-i)!\prod_{i<j}^a\sigma_i-\sigma_j, &\text{$b\geq a$},\\
\prod_{i=1}^b(b-i)!\prod_{i<j}^a\sigma_i-\sigma_j, & \text{$b<a$},
\end{cases}
\end{flalign}
and $\mathbf{G}$ is a $a\times(a-n)$ matrix, whose $(i,j)$th element is $\sigma_i^{j-1}$. $\bm{\sigma}=(\sigma_1,...\sigma_a)$ are the eigenvalues of $\mathbf{R}_a$, such that $\sigma_1>...>\sigma_a>0$. $\mathbf{E}(\bm{\lambda})$ is a $a\times n$ matrix, whose $(i,j)$th element is $[\sigma_i^{a-n-1}\exp(-\lambda_{j-a+n}/\sigma_i)]$. 

\vspace{0.125in}
Integrating Eqn. (\ref{Base1}) over $D_3$, we can get the probability $p$ as
\begin{flalign}\label{ProT11}
p&=K_0^{-1}\int_{D_3}\det[\mathbf{G},\mathbf{E}(\bm{\lambda})]\prod_{i<j}^n(\lambda_i-\lambda_j)\prod_{i=1}^n\lambda_i^{b-n}d\lambda_i.
\end{flalign}
Performing the Laplace expansion over the first $a-n$ columns of $[\mathbf{G}, \mathbf{E}(\bm{\lambda})]$, we gave
\begin{equation}\label{change}
\det[\mathbf{G}, \mathbf{E}(\bm{\lambda})]=\sum_{\bm{\kappa}\in \mathcal{Q}(i)}(-1)^{\sum_{i=1}^{a-n}(\kappa_i+i)}\det[\mathbf{G}^{\bm{\kappa}}]\det[\mathbf{E}^{\bm{\kappa}}(\bm{\lambda})],
\end{equation}
where $\mathcal{Q}(i)$ is a set of all permutations $(\kappa_1,...,\kappa_a)$ of the integers $(1,...,a)$, such that $(\kappa_1<\kappa_2<...<\kappa_{a-n})$ and $(\kappa_{a-n+1}<\kappa_{a-n+2}<...<\kappa_a)$. Hence, $\sum_{\bm{\kappa}\in\mathcal{Q}(i)}$ denotes the summation over two combinations $(\kappa_1<\kappa_2<...<\kappa_{a-n})$ and $(\kappa_{a-n+1}<\kappa_{a-n+2}<...<\kappa_a)$. $[\mathbf{E}^{\bm{\kappa}}(\bm{\lambda})]$ is a $n\times n$ matrix, i.e., $[\mathbf{E}^{\bm{\kappa}}(\bm{\lambda})]_{i,j}=\sigma^{a-n-1}_{\kappa_{a-n+i}} \exp(-\lambda_j/\sigma_{\kappa_{a-n+i}})$ for $i, j=1,...,n$. $[\mathbf{G}^{\bm{\kappa}}]$ is a $(a-n)\times (a-n)$ Vandermonde matrix, i.e., $[\mathbf{G}^{\bm{\kappa}}]_{i,j}=\sigma_{\kappa_{i}}^{j-1}$ for $i, j=1,...,a-n$. When $a=n$, we set $\det[\mathbf{G}^{\bm{\kappa}}]=1$.

\newcounter{mytempeqncnt}
\begin{figure*}[t]
\begin{flalign}\label{change1}
&\det\big[\mathbf{E}^{\bm{\kappa}}(\bm{\lambda})\big]\prod_{i<j}^n(\lambda_i-\lambda_j)=\prod_{i=1}^n\sigma_{\kappa_{a-n+i}}^{a-n-1} \sum^{\sim}_q\sum^{\sim}_\iota(-1)^{\text{per}(\iota_1,\dots,\iota_n)}\prod_{i=1}^n\lambda_{q_i}^{\iota_i-1}\exp(-\frac{\lambda_{q_i}}{\sigma_{\kappa_{a-n+i}}}).
\end{flalign}
\hrulefill
\vspace*{4pt}
\end{figure*}

Next, we prove Eqn. (\ref{change1}) for simplifying Eqn. (\ref{ProT11}). In Eqn. (\ref{change1}), $\sum^{\sim}_q$ denotes the summation over all permutations $(q_1,\dots,q_n)$ of $(1,\dots,n)$, $\sum^{\sim}_\iota$ is the summation over all permutations $(\iota_1,\dots,\iota_n)$ of $(1,\dots,n)$, and per($\iota_1,\dots,\iota_n$) is either 0 or 1, corresponding to even or odd value of the permutation $(\iota_1,\dots,\iota_n)$. Then, $p$ can be written as
\begin{flalign}
p&=K_0^{-1}\int_{D_3}\det[\mathbf{G},\mathbf{E}(\bm{\lambda})]\prod_{i<j}^n(\lambda_i-\lambda_j)\prod_{i=1}^n\lambda_i^{b-n}d\lambda_i \\ \notag
&=K_0^{-1}\sum_{\bm{\kappa}\in \mathcal{Q}(i)}(-1)^{\sum_{i=1}^{a-n}(\kappa_i+i)}\det[\mathbf{G}^{\bm{\kappa}}] \prod_{i=1}^n\sigma_{\kappa_{a-n+i}}^{a-n-1} \sum^{\sim}_q\sum^{\sim}_\iota \\ \notag
&\times (-1)^{\text{per}(\iota_1,\dots,\iota_n)}\int_{D_3}\prod_{i=1}^n\lambda_{q_i}^{\iota_i-1}\exp(-\frac{\lambda_{q_i}}{\sigma_{\kappa_{a-n+i}}})\prod_{i=1}^n\lambda_i^{b-n}d\lambda_{q_i}\\ \notag
&=K_0^{-1}\sum_{\bm{\mu}\in\mathcal{P}(k)} \sum_{\bm{\kappa}\in \mathcal{Q}(i)}(-1)^{\sum_{i=1}^{a-n}(\kappa_i+i)}\det[\mathbf{G}^{\bm{\kappa}}]\prod_{i=1}^n\sigma_{\kappa_{a-n+i}}^{a-n-1} \\ \notag
&\times \sum^{\sim}_\iota(-1)^{\text{per}(\iota_1,\dots,\iota_n)}I_1(\mu,\iota,\kappa)I_2(\mu,\iota,\kappa),
\end{flalign}
where $\sum_q^{\sim}=\sum_{\bm{\mu}\in\mathcal{P}(k)}\sum^{\sim}_{q_{\mu_{\psi}}}\sum^{\sim}_{q_{\mu_{\omega}}}$, and $\sum^{\sim}_{q_{\mu_{\psi}}}$ denotes the summation over the permutations $(q_{\mu_1},\dots,q_{\mu_{k-1}})$ of $(1,\dots,k-1)$, $\sum^{\sim}_{q_{\mu_{\omega}}}$ calculates the summation over the permutations $(q_{\mu_k},\dots,q_{\mu_n})$ of $(k,\dots,n)$, $\sum_{\bm{\mu}\in\mathcal{P}(k)}$ is the summation over the combination of sets $(\mu_1<\mu_2<\dots<\mu_{k-1})$ and $(\mu_k<\mu_{k+1}<\dots<\mu_n)$, and $(\mu_1,\dots,\mu_n)$ is a permutation of $(1,\dots,n)$. From \cite[Eqs. (4.20) and (4.21)]{Ratnarajah2003}, we obtain
\begin{flalign}
I_1(\mu,\iota,\kappa)&=\sum^{\sim}_{q_{\mu_{\psi}}}\int_{D_4}\prod_{i=1}^{k-1}\lambda_{q_{\mu_i}}^{b-n+\iota_{i}-1}\exp(-\frac{\lambda_{q_{\mu_i}}}{\sigma_{\kappa_{a-n+\mu_i}}})d\lambda_{q_{\mu_i}} \notag\\
&=\prod_{i=1}^{k-1}\int_x^{\infty}\lambda_{\mu_i}^{b-n+\iota_{i}-1}\exp(-\frac{\lambda_{\mu_i}}{\sigma_{\kappa_{a-n+\mu_i}}})d\lambda_{\mu_i} \notag\\
&=\prod_{i=1}^{k-1}\sigma_{\kappa_{a-n+\mu_i}}^{b-n+\iota_i}\Gamma(b-n+\iota_i,\frac{\lambda_{\mu_i}}{\sigma_{\kappa_{a-n+\mu_i}}}), \\ \notag
I_2(\mu,\iota,\kappa)&=\sum^{\sim}_{q_{\mu_{\omega}}}\int_{D_5}\prod_{i=k}^{n}\lambda_{q_{\mu_i}}^{b-n+\iota_{i}-1}\exp(-\frac{\lambda_{q_{\mu_i}}}{\sigma_{\kappa_{a-n+\mu_i}}})d\lambda_{q_{\mu_i}} \notag\\
&=\prod_{i=k}^{n}\int_0^x\lambda_{\mu_i}^{\iota_{i}-1}\exp(-\frac{\lambda_{\mu_i}}{\sigma_{\kappa_{a-n+\mu_i}}})d\lambda_{\mu_i} \notag\\
&=\prod_{i=k}^{n}\sigma_{\kappa_{a-n+\mu_i}}^{b-n+\iota_i}\gamma(b-n+\iota_i,\frac{\lambda_{\mu_i}}{\sigma_{\kappa_{a-n+\mu_i}}}), 
\end{flalign}
where $D_4=\{x<\lambda_{k-1}<\dots<\lambda_1<\infty\}$ and $D_5=\{0<\lambda_n<\dots<\lambda_k<x\}$. $\iota_i$ is the $i$th position after re-ordering $(\iota_1,\dots,\iota_n)$, which can be viewed as the column index of the determinant of an $(n\times n)$ matrix. $\mu_i$ is the row index of the determinant of the $(n\times n)$ matrix dependent on $k$. Hence, $\sum^{\sim}_\iota(-1)^{\text{per}(\iota_1,\dots,\iota_n)}I_1(\mu,\iota,\kappa)I_2(\mu,\iota,\kappa)$ denotes the determinant of a matrix, each element of which is expressed by $[\mathbf{\Theta}(\bm{\mu},\bm{\sigma},\bm{\kappa},k;x)]_{\mu_i,i}$. We can re-define the order index numbers of rows and columns of the determinant as $u$ and $\mu_v$. Finally, we get
\begin{flalign}\label{p2}
p&=K_0^{-1}\sum_{\bm{\mu}\in\mathcal{P}(k)} \sum_{\bm{\kappa}\in \mathcal{Q}(i)}(-1)^{\sum_{i=1}^{a-n}(\kappa_i+i)}\det[\mathbf{G}^{\bm{\kappa}}] \\ \notag
& \times  \prod_{i=1}^n\sigma_{\kappa_{a-n+i}}^{a-n-1}\det\big[\mathbf{\Theta}(\bm{\mu},\bm{\sigma},\bm{\kappa},k;x)\big],
\end{flalign}
where $(n\times n)$ real matrix $\mathbf{\Theta}(\bm{\mu},\bm{\sigma},\bm{\kappa},k;x)$ is defined as
\begin{flalign}\label{Matrix2}
&\big[\mathbf{\Theta}(\bm{\mu},\bm{\sigma},\bm{\kappa},k;x)\big]_{u,\mu_v} \\ \notag
&=\begin{cases}
\sigma_{\kappa_{a-n+u}}^{b-n+\mu_v}\Gamma(b-n+\mu_v,\frac{x}{\sigma_{\kappa_{a-n+u}}}),&\text{$v=1,...,k-1$},\\
\sigma_{\kappa_{a-n+u}}^{b-n+\mu_v}\gamma(b-n+\mu_v,\frac{x}{\sigma_{\kappa_{a-n+u}}}),&\text{$v=k,...,n$},
\end{cases}
\end{flalign}
for $u,v=1, ..., n$, where $\Gamma(\cdot,\cdot)$ and $\gamma(\cdot,\cdot)$ are the upper and lower incomplete Gamma functions defined in Eqns. (\ref{UGmma}) and (\ref{LGmma}). 

Since we have
\begin{flalign}\label{replace1}
&\prod_{i=1}^n\sigma_{\kappa_{a-n+i}}^{a-n-1}\det\big[\mathbf{\Theta}(\bm{\mu},\bm{\sigma},\bm{\kappa},k;x)\big]\\ \notag 
&= \prod_{i=1}^n\sigma_{\kappa_{a-n+i}}^{b-n}\det[\mathbf{\Psi}(\bm{\mu},\bm{\sigma},\bm{\kappa},k;x)],
\end{flalign}
where $(n\times n)$ real matrix $\mathbf{\Psi}(\bm{\mu},\bm{\sigma},k,\kappa;x)$ is defined as
\begin{flalign}\label{Matrix2}
&[\mathbf{\Psi}(\bm{\mu},\bm{\sigma},\bm{\kappa},k;x)]_{u,\mu_v}\\ \notag
&=\begin{cases}
\sigma_{\kappa_{a-n+u}}^{a-n+\mu_v-1}\Gamma(b-n+\mu_v,\frac{x}{\sigma_{\kappa_{a-n+u}}}),&\text{$v=1,...,k-1$},\\
\sigma_{\kappa_{a-n+u}}^{a-n+\mu_v-1}\gamma(b-n+\mu_v,\frac{x}{\sigma_{\kappa_{a-n+u}}}),&\text{$v=k,...,n$},
\end{cases}
\end{flalign}
for $u,v=1, ..., n$. Substituting Eqn. (\ref{replace1}) to Eqn. (\ref{p2}) and performing the inverse Laplace expansion of Eqn. (\ref{p2}), we obtain
\begin{flalign}\label{correcdf}
p&=K_0^{-1}\prod_{i=1}^n\sigma_{i}^{b-n}\sum_{\bm{\mu}\in\mathcal{P}(k)} \det[\mathbf{G},\mathbf{\Psi}(\bm{\mu},\bm{\sigma},k;x)] \\ \notag
&=K^{-1}\sum_{\bm{\mu}\in\mathcal{P}(k)} \det[\mathbf{G},\mathbf{\Psi}(\bm{\mu},\bm{\sigma},k;x)],
\end{flalign}
where 
\begin{equation}\label{k}
K=\prod_{i<j}^n\sigma_i-\sigma_j \prod_{i=1}^n(b-i)!.
\end{equation}

$F_{\lambda_k}(x)$ can be expressed by
\begin{flalign}\label{cdf0}
F_{\lambda_k}(x)=K^{-1}\sum_{i=1}^k\sum_{\bm{\mu}\in\mathcal{P}(i)}\det[\mathbf{G},\mathbf{\Psi}(\bm{\mu},\bm{\sigma},i;x)],
\end{flalign}
which is the marginal cdf of the $k$th largest eigenvalue $\lambda_k$ of a receiver-side correlated central Wishart matrix $\mathbf{W}$$\sim$$W_n(m,\mathbf{0}_n,\mathbf{R}_a)$. The marginal pdf of the $k$th largest eigenvalue can be easily derived from the derivative of a determinant as shown in \cite{Christiano1964}, which is
\begin{align}
f_{\lambda_k}(x)&=\frac{d}{dx}\bigg\{K^{-1}\sum_{i=1}^k\sum_{\bm{\mu}\in\mathcal{P}(i)}\det\big[\mathbf{G},\mathbf{\Psi}(\bm{\mu},\bm{\sigma},i;x)\big]\bigg\}\notag\\ 
&=K^{-1}\sum_{i=1}^k\sum_{\bm{\mu}\in\mathcal{P}(i)}\sum_{j=1}^{n}\det\big[\mathbf{G},\mathbf{\Omega}(\bm{\mu},\bm{\sigma},i,j;x)\big],
\end{align}
where $(n\times n)$ real matrix $\mathbf{\Omega}(\bm{\mu},\bm{\sigma},i,j;x)$ is defined in Eqn. (\ref{Matrixim}). This completes the proof.  \hfill $\blacksquare$

\subsection{Proof of Theorem 4}
According to Jensen's inequality, we have
\begin{flalign}
C_{\mathbf{A}}(\mathbf{R}_a, \rho,\eta)=&\sum_{i=1}^{\eta}\text{E}\big\{\log_2[1+(P/t)\lambda_i(\mathbf{A}\mathbf{A}^{\dagger})]\big\} \\ \notag
\leq & \sum_{i=1}^{\eta}\log_2\big\{1+(P/t)\text{E}[\lambda_i(\mathbf{A}\mathbf{A}^{\dagger})]\big\},
\end{flalign}
where $\lambda_1(\mathbf{A}\mathbf{A}^{\dagger})>\lambda_2(\mathbf{A}\mathbf{A}^{\dagger})>\cdots>\lambda_n(\mathbf{A}\mathbf{A}^{\dagger})$ are the ordered eigenvalues of $\mathbf{A}\mathbf{A}^{\dagger}$. Thus, $C_{\mathbf{H}}(\mathbf{R}_r, \rho,s_1)$ in Eqn. (\ref{esc}) can be expressed as
\begin{flalign}
C_{\mathbf{H}}(\mathbf{R}_r, \rho,s_1)=\chi_1=\sum_{i=1}^{s_1}\log_2\big\{1+\rho\text{E}[\lambda_i(\mathbf{H}\mathbf{H}^{\dagger})]\big\}.
\end{flalign}	

From \cite[Eqn. (21)]{Zhang2005} or \cite[Eqn. (27)]{Cui2005}, we get
\begin{flalign} \label{LA1}
C_{\mathbf{H}_3}(\mathbf{R}_e, \rho,n_1)=\log_2\bigg[1+\sum_{k=1}^{e}\rho^k\prod_{i=0}^{k-1}(m_1-i)\varrho_k\bigg],
\end{flalign} 
and
\begin{flalign} \label{LA1}
C_{\mathbf{H}_4}(\mathbf{R}_e,\rho,e)=\log_2\bigg[1+\sum_{k=1}^{e}\rho^k\prod_{i=0}^{k-1}(t-i)\varrho_k \bigg],
\end{flalign}
respectively, where $\varrho_k$, $n_1$, and $m_1$ are defined in Eqns. (\ref{varrho}) and (\ref{n1m1}). We can simplify $C_{\mathbf{H}_3}(\mathbf{R}_e, \rho,n_1)-C_{\mathbf{H}_4}(\mathbf{R}_e,\rho,e)$ as
\begin{flalign}
&C_{\mathbf{H}_3}(\mathbf{R}_e, \rho,n_1)-C_{\mathbf{H}_4}(\mathbf{R}_e,\rho,e)\\ \notag
&=\chi_2= \log_2\bigg[\frac{1+\sum_{k=1}^{e}\rho^k\prod_{i=0}^{k-1}(m_1-i)\varrho_k}{1+\sum_{k=1}^{e}\rho^k\prod_{i=0}^{k-1}(t-i)\varrho_k} \bigg].
\end{flalign}
Hence, Eqn. (\ref{lowesc}) can be expressed approximately by
\begin{flalign}
R^{\text{app}}_s&=[\chi_1+\chi_2]^{+}.
\end{flalign}
This completes the proof.  \hfill $\blacksquare$

\balance
\bibliographystyle{IEEEtran}
\bibliography{IEEEabrv}

\begin{thebibliography}{10}
\providecommand{\url}[1]{#1}
\csname url@samestyle\endcsname
\providecommand{\newblock}{\relax}
\providecommand{\bibinfo}[2]{#2}
\providecommand{\BIBentrySTDinterwordspacing}{\spaceskip=0pt\relax}
\providecommand{\BIBentryALTinterwordstretchfactor}{4}
\providecommand{\BIBentryALTinterwordspacing}{\spaceskip=\fontdimen2\font plus
\BIBentryALTinterwordstretchfactor\fontdimen3\font minus
  \fontdimen4\font\relax}
\providecommand{\BIBforeignlanguage}[2]{{%
\expandafter\ifx\csname l@#1\endcsname\relax
\typeout{** WARNING: IEEEtran.bst: No hyphenation pattern has been}%
\typeout{** loaded for the language `#1'. Using the pattern for}%
\typeout{** the default language instead.}%
\else
\language=\csname l@#1\endcsname
\fi
#2}}
\providecommand{\BIBdecl}{\relax}
\BIBdecl

\bibitem{WangHeter2018}
W.~{Wang}, K.~C. {Teh}, S.~{Luo}, and K.~H. {Li}, ``Physical layer security in
  heterogeneous networks with pilot attack: {A} stochastic geometry approach,''
  \emph{IEEE Trans. Commun.}, vol.~66, no.~12, pp. 6437--6449, Dec. 2018.

\bibitem{Oggier2008}
F.~Oggier and B.~Hassibi, ``The secrecy capacity of the {MIMO} wiretap
  channel,'' in \emph{Proc. IEEE International Symposium on Information
  Theory}, Jul. 2008, pp. 524--528.

\bibitem{LiuMIMO2017}
Y.~Liu, Z.~Qin, M.~Elkashlan, Y.~Gao, and L.~Hanzo, ``Enhancing the physical
  layer security of non-orthogonal multiple access in large-scale networks,''
  \emph{IEEE Trans. Wireless Commun.}, vol.~16, no.~3, pp. 1656--1672, Mar.
  2017.

\bibitem{Li2016}
B.~Li, Z.~Fei, and H.~Chen, ``Robust artificial noise-aided secure beamforming
  in wireless-powered non-regenerative relay networks,'' \emph{IEEE Access},
  vol.~4, pp. 7921--7929, 2016.

\bibitem{Yu2018sj}
H.~{Yu}, S.~{Guo}, Y.~{Yang}, and B.~{Xiao}, ``Optimal target secrecy rate and
  power allocation policy for a {SWIPT} system over a fading wiretap channel,''
  \emph{IEEE IEEE Syst. J.}, vol.~12, no.~4, pp. 3291--3302, Dec. 2018.

\bibitem{Negi05}
R.~Negi and S.~Goel, ``Secret communication using artificial noise,'' in
  \emph{Proc. IEEE Vehicular Technology Conference}, vol.~3, Sep. 2005, pp.
  1906--1910.

\bibitem{Goel2008}
S.~{Goel} and R.~{Negi}, ``Guaranteeing secrecy using artificial noise,''
  \emph{IEEE Tran. Wireless Commun.}, vol.~7, no.~6, pp. 2180--2189, Jun. 2008.

\bibitem{Tsai2014}
S.-H. Tsai and H.~V. Poor, ``Power allocation for artificial-noise secure
  {MIMO} precoding systems,'' \emph{IEEE Trans. Signal Process.}, vol.~62,
  no.~13, pp. 3479--3493, Jul. 2014.

\bibitem{Liu2015}
S.~{Liu}, Y.~{Hong}, and E.~{Viterbo}, ``Artificial noise revisited,''
  \emph{IEEE Trans. Inf. Theory}, vol.~61, no.~7, pp. 3901--3911, Jul. 2015.

\bibitem{Yun2017conf}
S.~{Yun}, J.~{Park}, S.~{Im}, and J.~{Ha}, ``On the secrecy rate of artificial
  noise assisted {MIMOME} channels with full-duplex receiver,'' in \emph{Proc.
  IEEE Wireless Communications and Networking Conference (WCNC)}, Mar. 2017,
  pp. 1--6.

\bibitem{Yun2018}
S.~{Yun}, S.~{Im}, I.~{Kim}, and J.~{Ha}, ``On the secrecy rate and optimal
  power allocation for artificial noise assisted {MIMOME} channels,''
  \emph{IEEE Trans. Veh. Technol.}, vol.~67, no.~4, pp. 3098--3113, Apr. 2018.

\bibitem{Wang2017AN}
W.~Wang, K.~C. Teh, and K.~H. Li, ``Artificial noise aided physical layer
  security in multi-antenna small-cell networks,'' \emph{IEEE Trans. Inf.
  Forensics Security}, vol.~12, no.~6, pp. 1470--1482, Jun. 2017.

\bibitem{Shu2018}
F.~{Shu}, W.~{Zhu}, X.~{Zhou}, J.~{Li}, and J.~{Lu}, ``Robust secure
  transmission of using main-lobe-integration-based leakage beamforming in
  directional modulation {MU-MIMO} systems,'' \emph{IEEE Syst. J.}, vol.~12,
  no.~4, pp. 3775--3785, Dec. 2018.

\bibitem{Liuwishart2017}
Y.~Liu, H.~H. Chen, and L.~Wang, ``Secrecy capacity analysis of artificial
  noisy {MIMO} channels--{An} approach based on ordered eigenvalues of
  {Wishart} matrices,'' \emph{IEEE Trans. Inf. Forensics Security}, vol.~12,
  no.~3, pp. 617--630, Mar. 2017.

\bibitem{Ahmed2018}
M.~{Ahmed} and L.~{Bai}, ``Secrecy capacity of artificial noise aided secure
  communication in {MIMO} {Rician} channels,'' \emph{IEEE Access}, vol.~6, pp.
  7921--7929, 2018.

\bibitem{Zheng2018}
Z.~{Zheng}, Z.~J. {Haas}, and M.~{Kieburg}, ``Secrecy rate of cooperative
  {MIMO} in the presence of a location constrained eavesdropper,'' \emph{IEEE
  Tran. Commun.}, vol.~67, no.~2, pp. 1356--1370, Feb. 2019.

\bibitem{3gpp}
\emph{Spatial channel model for Multiple Input Multiple Output ({MIMO})
  simulations}, 3rd Generation Partnership Project Std. 3GPP TR 25.996, 9 2003,
  rev. 6.

\bibitem{Simon2006}
S.~H. Simon, A.~L. Moustakas, and L.~Marinelli, ``Capacity and character
  expansions: {Moment-Generating} function and other exact results for {MIMO}
  correlated channels,'' \emph{IEEE Trans. Inf. Theory}, vol.~52, no.~12, pp.
  5336--5351, Dec. 2006.

\bibitem{Bolcskei2003}
H.~Bolcskei, M.~Borgmann, and A.~J. Paulraj, ``Impact of the propagation
  environment on the performance of space-frequency coded {MIMO-OFDM},''
  \emph{IEEE J. Sel. Areas Commun.}, vol.~21, no.~3, pp. 427--439, Apr. 2003.

\bibitem{Li2018}
B.~Li, Z.~Fei, Z.~Chu, F.~Zhou, K.~Wong, and P.~Xiao, ``Robust
  chance-constrained secure transmission for cognitive satellite–terrestrial
  networks,'' \emph{IEEE Trans. Veh. Technol.}, vol.~67, no.~5, pp. 4208--4219,
  May 2018.

\bibitem{Zorgui2016}
M.~Zorgui, Z.~Rezki, B.~Alomair, E.~A. Jorswieck, and M.~Alouini, ``On the
  ergodic secret-key agreement over spatially correlated multiple-antenna
  channels with public discussion,'' \emph{IEEE Trans. Signal Process.},
  vol.~64, no.~2, pp. 495--510, Jan. 2016.

\bibitem{Zhang2016}
J.~Zhang, C.~Yuen, C.~Wen, S.~Jin, K.~Wong, and H.~Zhu, ``Large system secrecy
  rate analysis for {SWIPT MIMO} wiretap channels,'' \emph{IEEE Trans. Inf.
  Forensics Security}, vol.~11, no.~1, pp. 74--85, Jan. 2016.

\bibitem{Lyu2018}
F.~{Lyu}, N.~{Cheng}, H.~{Zhu}, H.~{Zhou}, W.~{Xu}, M.~{Li}, and X.~{Shen},
  ``Intelligent context-aware communication paradigm design for {IoVs} based on
  data analytics,'' \emph{IEEE Netw.}, vol.~32, no.~6, pp. 74--82, Nov. 2018.

\bibitem{Ordonez2009}
L.~G. {Ordonez}, D.~P. {Palomar}, and J.~R. {Fonollosa}, ``Ordered eigenvalues
  of a general class of {Hermitian} random matrices with application to the
  performance analysis of {MIMO} systems,'' \emph{IEEE Trans. Signal Process.},
  vol.~57, no.~2, pp. 672--689, Feb. 2009.

\bibitem{Ratnarajah2003}
T.~Ratnarajah, ``Topics in complex random matrices and information theory,''
  Master's thesis, Mathematics and Statistics of University of Ottawa, May
  2003.

\bibitem{McKay2005}
M.~R. McKay and I.~B. Collings, ``General capacity bounds for spatially
  correlated {Rician MIMO} channels,'' \emph{IEEE Trans. Inf. Theory}, vol.~51,
  no.~9, pp. 3121--3145, Sep. 2005.

\bibitem{Kobayashi2011}
M.~{Kobayashi}, N.~{Jindal}, and G.~{Caire}, ``Training and feedback
  optimization for multiuser {MIMO} downlink,'' \emph{IEEE Trans. Commun.},
  vol.~59, no.~8, pp. 2228--2240, Aug. 2011.

\bibitem{Liuma2015}
T.~{Liu}, P.~{Lin}, S.~{Lin}, Y.~P. {Hong}, and E.~A. {Jorswieck}, ``To avoid
  or not to avoid {CSI} leakage in physical layer secret communication
  systems,'' \emph{IEEE Commun. Magazine}, vol.~53, no.~12, pp. 19--25, Dec.
  2015.

\bibitem{tse2005fundamentals}
D.~Tse and P.~Viswanath, \emph{Fundamentals of wireless communication}.\hskip
  1em plus 0.5em minus 0.4em\relax Cambridge, England: Cambridge university
  press, 2005.

\bibitem{wishart1928generalised}
J.~Wishart, ``The generalised product moment distribution in samples from a
  normal multivariate population,'' \emph{Biometrika}, pp. 32--52, Jul. 1928.

\bibitem{horn2012matrix}
R.~A. Horn and C.~R. Johnson, \emph{Matrix analysis}.\hskip 1em plus 0.5em
  minus 0.4em\relax Cambridge, England: Cambridge university press, 2012.

\bibitem{Zhang2005}
Q.~T. {Zhang}, X.~W. {Cui}, and X.~M. {Li}, ``Very tight capacity bounds for
  {MIMO}-correlated {Rayleigh}-fading channels,'' \emph{IEEE Trans. Wireless
  Commun.}, vol.~4, no.~2, pp. 681--688, Mar. 2005.

\bibitem{Cui2005}
X.~W. {Cui}, Q.~T. {Zhang}, and Z.~M. {Feng}, ``Generic procedure for tightly
  bounding the capacity of {MIMO} correlated {Rician} fading channels,''
  \emph{IEEE Trans. Commun.}, vol.~53, no.~5, pp. 890--898, May 2005.

\bibitem{baik2006eigenvalues}
J.~Baik and J.~W. Silverstein, ``Eigenvalues of large sample covariance
  matrices of spiked population models,'' \emph{Journal of multivariate
  analysis}, vol.~97, no.~6, pp. 1382--1408, Jul. 2006.

\bibitem{Ghaderipoor2012}
A.~Ghaderipoor, C.~Tellambura, and A.~Paulraj, ``On the application of
  character expansions for {MIMO} capacity analysis,'' \emph{IEEE Trans. Inf.
  Theory}, vol.~58, no.~5, pp. 2950--2962, May 2012.

\bibitem{HORN1992337}
R.~A. Horn and R.~Mathias, ``Block-matrix generalizations of {Schur's} basic
  theorems on {Hadamard} products,'' \emph{Linear Algebra and its
  Applications}, vol. 172, pp. 337--346, Jul. 1992.

\bibitem{BAPAT1985107}
R.~Bapat and V.~Sunder, ``On majorization and schur products,'' \emph{Linear
  Algebra and its Applications}, vol.~72, pp. 107 -- 117, Dec. 1985.

\bibitem{gupta1999matrix}
A.~K. Gupta and D.~K. Nagar, \emph{Matrix variate distributions}.\hskip 1em
  plus 0.5em minus 0.4em\relax Boca Raton, Florida: CRC Press, 2018.

\bibitem{James1964}
A.~T. James \emph{et~al.}, ``Distributions of matrix variates and latent roots
  derived from normal samples,'' \emph{The Annals of Mathematical Statistics},
  vol.~35, no.~2, pp. 475--501, 1964.

\bibitem{Christiano1964}
\BIBentryALTinterwordspacing
J.~E.~H. John G.~Christiano, ``On the $n$-th derivative of a determinant of the
  $j$-th order,'' \emph{Math. Magazine}, vol.~37, no.~4, pp. 215--217, Sep.
  1964. [Online]. Available: \url{http://www.jstor.org/stable/2688589}
\BIBentrySTDinterwordspacing

\end{thebibliography}

\begin{IEEEbiography}[{\includegraphics[width=1in,height=1.25in,clip,keepaspectratio]{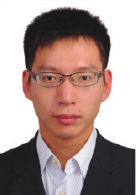}}]{Yi-liang Liu} (S'18) received the B.E and M.Sc degrees in Computer Science and Communication Engineering from Jiangsu University, Zhenjiang, China, in 2012 and 2015, respectively. He was a Visiting Research Student with the Department of Engineering Science, National Cheng Kung University, Tainan, Taiwan, from 2014 to 2015. He is currently working toward his PhD degree in the Communication Research Centre, Harbin Institute of Technology, China. His research interests include security of wireless communications, physical layer security, and intelligent connected vehicles.
\end{IEEEbiography}
\begin{IEEEbiography}[{\includegraphics[width=1in,height=1.25in,clip,keepaspectratio]{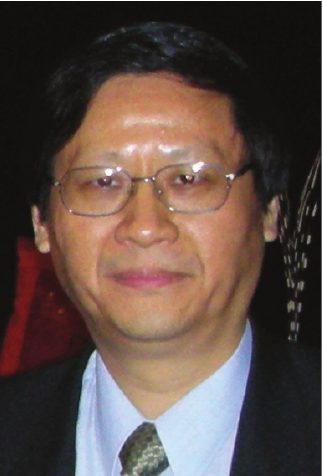}}]{Hsiao-Hwa Chen} (S'89-M'91-SM'00-F'10) received the B.Sc. and M.Sc. degrees from Zhejiang University, China, in 1982, and the Ph.D. degree from the University of Oulu, Finland, in 1985 and 1991, respectively. He is currently a Distinguished Professor with the Department of Engineering Science, National Cheng Kung University, Taiwan. He is a Fellow of the IET. He was a Recipient of the Best Paper Award at IEEE WCNC 2008 and the IEEE 2016 Jack Neubauer Memorial Award. He served as the general chair, the TPC chair, and the symposium chair for major international conferences. He served or serves as an editor or a guest editor for numerous technical journals. He is the Founding Editor-in-Chief of Security and Communication Networks Journal (Wiley). He served as the Editor-in-Chief for IEEE Wireless Communications from 2012 to 2015. He was an Elected Member-at-Large of IEEE ComSoc from 2015 to 2016.
\end{IEEEbiography}
\begin{IEEEbiography}[{\includegraphics[width=1in,height=1.25in,clip,keepaspectratio]{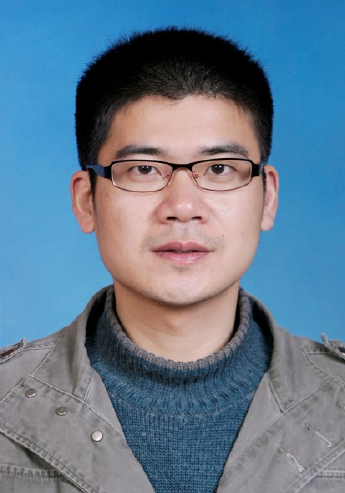}}]{Liang-min WANG} (M'12) received the BS degree in computational mathematics in Jilin University, Changchun, China, in 1999, and the PhD degree in cryptology from Xidian University, Xi'an, China, in 2007. He is a full professor with the School of Computer Science and Communication Engineering, Jiangsu University, Zhenjiang, China. He has been honored as a ``Wan-Jiang Scholar'' of Anhui Province since Nov. 2013. Now his research interests include data security \& privacy. He has published more than 60 technical papers at premium international journals and conferences, like the IEEE Transactions on Intelligent Transportation Systems, the IEEE Transactions on Vehicular Technology, IEEE Global Communications Conference, IEEE Wireless Communications and Networking Conference. He has served as a TPC member of many IEEE conferences, such as IEEE ICC, IEEE HPCC, IEEE TrustCOM. Now he is an associate editor of Security and Communication Networks, a member of the IEEE, ACM, and a senior member of Chinese Computer Federation.
\end{IEEEbiography}
\begin{IEEEbiography}[{\includegraphics[width=1in,height=1.25in,clip,keepaspectratio]{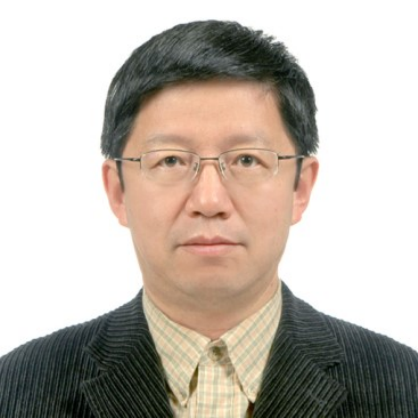}}]{Weixiao Meng} (M'04–SM'10) received the B.Eng., M.Eng., and Ph.D. degrees from the Harbin Institute of Technology (HIT), Harbin, China, in 1990, 1995, and 2000, respectively. From 1998 to 1999, he worked at NTT DoCoMo on adaptive array antennas and dynamic resource allocation for beyond 3G as a senior visiting researcher. He is now a full professor and the vice dean of the School of Electronics and Information Engineering of HIT. His research interests include
broadband wireless communications and networking, MIMO, GNSS receivers and wireless localization technologies. He has published three books and over 220 papers in journals and international conferences. He is the chair of the IEEE Communications Society Harbin Chapter, a Fellow of the China Institute of Electronics, and a senior member of the IEEE and the China Institute of Communication.
\end{IEEEbiography}

\end{document}